\documentclass[a4paper,11pt]{article}
\pdfoutput=1 

\usepackage{graphicx}  
\usepackage{dcolumn}   
\usepackage{bm,relsize}        
\usepackage{amssymb, amsmath}
\usepackage{textcomp}
\usepackage{wasysym}
\usepackage{slashed}
\usepackage{caption, subcaption}
\usepackage{multirow}
\usepackage{gensymb}

\usepackage{lipsum, color}
\usepackage[usenames,dvipsnames,svgnames]{xcolor}
\usepackage{braket}
\usepackage{jheppub} 
\usepackage{booktabs}
\usepackage{pdflscape}
\usepackage[utf8]{inputenc} 

\def\beq{\begin{equation}}
\def\eeq{\end{equation}}
 \def\be{\begin{equation}} \def\ee{\end{equation}}
\def\bea{\begin{eqnarray}} \def\eea{\end{eqnarray}}

\newcommand{\g}{\gamma}

\newcommand{\bi}{\begin{itemize} }
\newcommand{\ei}{\end{itemize} }
\newcommand{\lp}{\left(}
\newcommand{\rp}{\right)}

\newcommand{\e}{e}

\usepackage{caption}
\usepackage{subcaption}
\usepackage{multirow}
\usepackage[normalem]{ulem}
\allowdisplaybreaks

\usepackage[capitalise, english]{cleveref}

\begin{document}

\title{Searching for neutrino polarizability at DUNE}

\author[a]{Sam Carey,}
\author[b]{Pedro Machado,}
\author[a]{Gil Paz,}
\author[c]{Alexey A. Petrov,}
\author[d]{Alexandre Sousa,}
\author[e,f]{Michele Tammaro}
\author[d]{and Jure Zupan}
\affiliation[a]{Department of Physics and Astronomy, Wayne State University, Detroit, Michigan 48201, USA}
\affiliation[b]{Particle Theory Department, Fermilab, P.O. Box 500, Batavia, IL 60510, USA}
\affiliation[c]{Department of Physics and Astronomy, University of South Carolina, Columbia, South Carolina 29208, USA}
\affiliation[d]{Department of Physics, University of Cincinnati, Cincinnati, Ohio 45221, USA}
\affiliation[e]{INFN Sezione di Firenze, Via G. Sansone 1, I-50019 Sesto Fiorentino, Italy}
\affiliation[f]{Galileo Galilei Institute for Theoretical Physics, Largo Enrico Fermi 2, I-50125 Firenze, Italy}

\emailAdd{samcarey@wayne.edu}
\emailAdd{pmachado@fnal.gov}
\emailAdd{gilpaz@wayne.edu}
\emailAdd{apetrov@sc.edu}
\emailAdd{alex.sousa@uc.edu}
\emailAdd{michele.tammaro@fi.infn.it}
\emailAdd{zupanje@ucmail.uc.edu}

\date{\today}

\preprint{FERMILAB-PUB-25-0472-T,USC-TH-2025-02, WSU-HEP-2501}

\abstract{
We perform an initial study of DUNE's sensitivity to enhanced neutrino polarizability within models of light scalar mediators.  
We identify two possible signatures of polarizability due to neutrino scattering on electrons and due to coherent scattering on argon nuclei. These result in either one or two separated electromagnetic showers, respectively, with no associated hadronic activity.
For each signature we compute the signal rates and the relevant backgrounds, obtaining the projected reach of the DUNE near detector. We then compare this with the current astrophysical and terrestrial bounds on light scalar models coupling to neutrinos and/or photons.}

\maketitle

\section{Introduction}
\label{sec:Introduction}
 
Electromagnetic interactions of neutrinos offer a powerful probe of various New Physics (NP) scenarios \cite{Giunti:2024gec}.
Since neutrinos are electrically neutral, their interactions with photons are highly suppressed in the Standard Model (SM). These interactions only occur through higher-dimensional operators, such as the dimension-5 magnetic moment operators $(1/2)\lambda_{\nu,{ij}}  \big(\bar \nu_i^c \sigma^{\mu\nu} P_L\nu_j\big) F_{\mu\nu}$ or the dimension-7 neutrino polarizability operators $(1/2)\alpha_{\nu, ij} \big(\bar \nu_i^c  P_L\nu_j\big) F_{\mu\nu} F^{\mu\nu}$ and $(1/2)\tilde \alpha_{\nu, ij} \big(\bar \nu_i^c  P_L\nu_j\big) F_{\mu\nu} \tilde F^{\mu\nu}$ \cite{Bansal:2022zpi}. In this paper we focus on the interactions that generate the polarizability of neutrinos. 

Neutrino magnetic moments and neutrino polarizability differ in both search strategies and the types of NP that can generate large effects. For instance, detector-based searches for anomalously large neutrino magnetic moments rely on the fact that magnetic moment-induced neutrino scattering is enhanced at low recoil energies. Therefore, low-threshold searches using solar neutrinos, such as those performed at Borexino and XENON-1T, are especially important~\cite{BOREXINO:2018ohr, XENON:2020rca}. Although low-threshold experiments using solar neutrinos can also be important for neutrino polarizability searches, we demonstrate in this paper that searches utilizing higher-energy neutrinos in DUNE have a clear advantage in certain regions of the UV parameter space.  

In the SM, the neutrino magnetic moment and polarizability appear at one loop and are suppressed by the ratio of neutrino masses to the electroweak scale, giving $\lambda_{\nu, ii}\sim {\mathcal O}( e m_{\nu_i}/16 \pi^2 v^2) \sim 10^{-18} (m_{\nu_i}/\text{eV})\mu_B$ \cite{Giunti:2014ixa,Giunti:2024gec}, and $\alpha_{\nu, ii}\sim {\mathcal O}( e^2 m_{\nu_i}/16 \pi^2 v^2) $, where $m_{\nu_i}$ is the neutrino mass and $\mu_B$ is the Bohr magneton. The off-diagonal terms are further suppressed by the charged lepton masses, $m_\ell^2/v^2$. In extensions of the SM both the neutrino magnetic moment and the neutrino polarizability can be enhanced, though for different reasons. In general, the magnetic moments need not be proportional to neutrino masses, since $\bar \nu_i^c  \sigma^{\mu \nu} P_L \nu_j$ is antisymmetric under the exchange of flavor indices, while the neutrino mass term $\bar \nu_i^c  \nu_j$ is symmetric. Explicit realizations of this so-called  {\it Voloshin mechanism} \cite{Voloshin:1987qy} can be found in Refs. \cite{Babu:1989wn,Babu:2020ivd,Babu:1990wv,Leurer:1989hx}. 
 
No such symmetry distinguishes the neutrino mass term and the neutrino polarizability operator, which implies that  polarizability is inevitably suppressed by tiny neutrino mass, $\alpha_{\nu, ii}\propto m_{\nu_i}$. However, as pointed out in Ref. \cite{Bansal:2022zpi}, polarizability can still be parametrically enhanced, if it is due to a tree-level exchange of a light spin-0 mediator, $\phi$, that couples to both photons (with coupling $\propto 1/\Lambda_\gamma$) and neutrinos (with coupling $\propto m_\nu/f_\phi$). 
At energies below the $\phi$ mass, $m_\phi$, this results in higher dimension operator with $\alpha_{\nu, ii} \propto m_{\nu,ii}/(m_\phi^2 f_\phi \Lambda_\gamma)$, so that the small mass $m_\phi$ at least partially compensates the suppression by the neutrino mass. 

For energies above $m_\phi$ the above EFT description is no longer valid. For instance, the mediator $\phi$ can now be produced on-shell, although this might not be the most effective way to search for models of enhanced neutrino polarizability. In this paper, for instance, we demonstrate that neutrino scattering via an off-shell exchange of the spin-0 mediator can produce distinct signatures at DUNE: a neutrino scattering event, whether on an electron or a nucleus, accompanied by the emission of a hard photon. 

The paper is organized as follows. In \cref{sec:NeutrinoPolarizability} we introduce the model of neutrino polarizability mediated by a light scalar, followed by the computation of the neutrino scattering rates on electrons and argon nuclei in \cref{sec:Scattering}. In \cref{sec:Signal&Background} we describe the signatures of neutrino polarizability at DUNE near detector (ND), identify the relevant sources of background and describe the simulation procedure. \Cref{sec:reach} contains the estimates of the expected reach at DUNE-ND, and its comparison with the existing constraints.  Conclusions are given in \cref{sec:Conclusions}. In \cref{app:OtherFF} we discuss the effects of  alternative versions of the nuclear form factor, and in \cref{sec:app:S:to:B} we present further distributions quantifying the signal-to-background ratios.  

\section{Scalar mediator induced neutrino polarizability}
\label{sec:NeutrinoPolarizability}

The neutrino polarizability operators are\footnote{For a list of higher dimension operators involving neutrino currents, up to dimension 7, see Ref. \cite{Altmannshofer:2018xyo}, and also Ref. \cite{Bansal:2022zpi} for a discussion focused on the electromagnetic neutrino interactions. In the notation of \cite{Altmannshofer:2018xyo,Bansal:2022zpi}, $\alpha_{\nu, ij}=\alpha_{\rm em} {\cal C}_{1,ij}^{(7)}/(12\pi \Lambda^3)$ and $\tilde \alpha_{\nu, ij}=\alpha_{\rm em} {\cal C}_{2,ij}^{(7)}/(8\pi \Lambda^3)$, with $\alpha_{\rm em}$ the fine structure constant, $\Lambda$ the new physics scale, and ${\cal C}_i$ the dimensionless Wilson coefficients. Ref.~\cite{Bansal:2022zpi} also defined electric ($\alpha_{E1,i}$) and magnetic $(\beta_{M1,i})$ scalar neutrino polarizabilities following conventions for nucleon polarizabilities, so that for non-relativistic neutrinos, ${\cal L}_{\rm NR}=2\pi \big(\alpha_{E1,i} \vec E^2+\beta_{M1,i} \vec B^2\big)\otimes 1_{\nu_i}$. In terms of neutrino polarizabilities in \eqref{eq:L:polariz}, we have $\alpha_{E1,i}=\beta_{M1,i}=\alpha_{\nu,ii}/(2\pi)$.
}
\beq
\label{eq:L:polariz}
{\cal L}_{\rm pol.}=\frac{1}{2}\alpha_{\nu, ij} \big(\bar \nu_i^c  P_L\nu_j\big) F_{\mu\nu} F^{\mu\nu}+\frac{1}{2} \tilde \alpha_{\nu, ij} \big(\bar \nu_i^c  P_L\nu_j\big) F_{\mu\nu} \tilde F^{\mu\nu},  
\eeq
where a summation over generation indices, $i,j=1,2,3,$ is understood. The neutrinos are assumed to be Majorana, so that $\nu_i^c=\nu_i$ (we use a four-component notation, following conventions from Ref. \cite{Dreiner:2008tw}). The neutrino polarizabilities $\alpha_{\nu, ij} $ and $\tilde \alpha_{\nu, ij}$ have dimensions of $\text{GeV}^{-3}$.  In any realistic UV model, their values are loop-suppressed by a factor ${\mathcal O}(\alpha_{\rm em}/4\pi)$.  In this work, we focus on the CP violating polarizability couplings, $\tilde \alpha_{\nu, ij}$, but most of the results for DUNE searches should also apply, up to ${\mathcal O}(1)$ corrections, to CP-conserving polarizabilities $\alpha_{\nu, ij}$. 

As pointed out in Ref. \cite{Bansal:2022zpi} a tree-level exchange of a light (pseudo-)scalar $\phi$ that couples to both neutrinos and photons induces an enhanced neutrino polarizability. That is, for energies below the mass of $\phi$, $m_\phi$, the pseudoscalar can be integrated out, so that the interaction Lagrangian
\beq\label{eq:Rayleigh:Lagrangian}
{\cal L}_{\rm int} =  - \frac{g_\g}4 \phi F_{\mu\nu}\tilde F^{\mu\nu} +\frac{1}{2}c_\nu^{ij} \phi \bar\nu_i^c P_L \nu_j +{\rm h.c.}\,, 
\eeq
leads to the dimension 7 CP-odd  polarizability operator in \eqref{eq:L:polariz} with 
\beq
\label{eq:tilde:alpha:nu}
\tilde \alpha_{\nu, ij}= c_\nu^{ij} \frac{g_\g}{4m_\phi^2}.
\eeq
Note that the $g_\g$ coupling in \eqref{eq:L:polariz} has dimensions $\text{GeV}^{-1}$, while $c_\nu^{ij}$ are dimensionless. A naive dimensional analysis estimate for the coupling to photons is $g_\g\sim {\mathcal O}(\alpha_{\rm em}/2\pi f_\phi=1/\Lambda_\gamma)$, where $f_\phi$ is the UV scale at which the couplings of $\phi$ to photons are generated, while the couplings to neutrinos are $c_\nu\sim {\mathcal O}(m_\nu/f_\phi)$, where for simplicity we assumed a single UV scale. For light $\phi$, i.e., for $m_\phi\ll f_\phi$, the neutrino polarizability is parametrically enhanced; while a naive guess for neutrino polarizability generated at scale $f_\phi$ would have been $\tilde \alpha_\nu \sim {\mathcal O}(\alpha_{\rm em} c_\nu/ 2\pi f_\phi^3)$, the small mass of $\phi$ enhances it to $\tilde \alpha_\nu \sim {\mathcal O}(\alpha_{\rm em} c_\nu / 2\pi f_\phi m_\phi^2)$.

The main goal of this paper is to estimate DUNE's sensitivity to neutrino polarizability, $\tilde \alpha_\nu$, \cref{eq:L:polariz}, with a specific focus on the enhanced neutrino polarizability generated from the tree-level exchange of a pseudo-scalar mediator, \cref{eq:Rayleigh:Lagrangian}. For neutrino scattering, see \cref{fig:Rayleigh:diagram}, 
three separate energy regimes can be identified, depending on the  size of $|q_\phi^2|$, the four-momentum squared carried by the mediator, compared to the mediator mass squared,  $m_\phi^2$. In the first regime $|q_\phi^2| \ll m_\phi^2$, and the pseudoscalar $\phi$ can be integrated out, resulting in an EFT description in \cref{eq:L:polariz} with $\alpha_{\nu,ij}=0$, and a nonzero $\tilde \alpha_{\nu, ij}$ given in \cref{eq:tilde:alpha:nu}. In the second regime, $|q_\phi^2| \sim  m_\phi^2$, and one needs to keep $\phi$ as a dynamical degree of freedom. The neutrino scattering cross section is then controlled by the interplay between the size of the momentum exchange and the mass of the mediator.
In the third regime, for very high energies, $|q_\phi^2| \gg m_\phi^2$, the mediator is effectively massless and thus the neutrino scattering cross section is proportional to $\sigma \propto 1/q_\phi^2$. 

The intermediate and the high energy regimes will be the ones relevant  for  the neutrino polarizability searches at DUNE-ND. Furthermore, in the numerical analysis, we will assume universal couplings to neutrinos
\beq
\label{eq:tilde:alpha:nu:uni}
\tilde \alpha_{\nu, ij}=\tilde \alpha_{\nu} \delta_{ij},
\eeq
though this choice can easily be relaxed (in that case the projected sensitivities to $\tilde \alpha_{\nu}$ are replaced by the appropriately weighted quadratic sums of $\tilde \alpha_{\nu, ij}$).

\section{Scattering processes in liquid argon from neutrino polarizability}
\label{sec:Scattering}

\begin{figure}[t]
	\centering
	\includegraphics[width=0.45\linewidth]{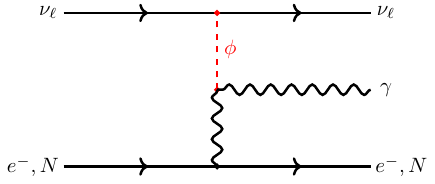}
    \hspace{1cm}
    \includegraphics[width=0.45\linewidth]{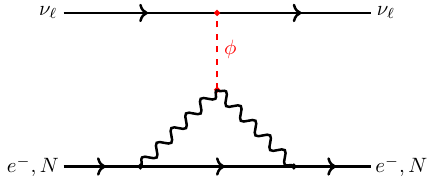}
	\caption{Feynman diagrams for the neutrino scattering on electron ($e^-$) or nucleus ($N$) at tree-level (left) and one-loop (right), mediated by a pseudoscalar particle $\phi$.}
	\label{fig:Rayleigh:diagram}
\end{figure}

A tree level exchange of a light scalar $\phi$ that couples to both photons and neutrinos induces two types of $2\to 3$ neutrino scatterings in the  DUNE-ND, see \cref{fig:Rayleigh:diagram} (left),
\beq
\label{eq:processes}
\nu_i e\to \nu_j \gamma e, \qquad \text{or}\qquad \nu_i N\to \nu_j \gamma N,
\eeq
where $e$ is an electron, and $N$ the argon nucleus. 
As we show below, the energetic final-state photon provides a useful handle for distinguishing the signal from the background.

Attaching both photon lines to the constituents of the argon atom, either electrons or nucleus, gives a one-loop induced correction to neutrino scattering on argon, see \cref{fig:Rayleigh:diagram} (right),
\beq
\label{eq:loop:process}
\nu_i +\text{Ar}\to \nu_j +\text{Ar}\,.
\eeq
Using naive dimensional analysis, we can estimate that the cross sections for $2\to 3$ scattering in \cref{eq:processes} scale as $\sigma \propto Z^2 c_\nu^2 g_\gamma^2 \e^2 /16\pi^2$ and the scattering rate for the loop induced process in \cref{eq:loop:process} is $\sigma \propto Z^4 c_\nu^2 g_\gamma^2 \e^4 /(16\pi^2)^2$, where $e$ is the electromagnetic coupling.
For argon, the atomic number is $Z=18$, and $(Z \e)^2=4\pi\alpha_{\rm em} Z^2\ll 16\pi^2$.  For neutrino scattering on argon, we therefore expect the $2\to 3$ scattering in \cref{eq:processes} to dominate and neglect the loop-induced corrections, \cref{eq:loop:process}. 

Next, we consider in \cref{sec:ScattOnElectrons} and  \cref{sec:ScattOnargon} the two signal processes in \cref{eq:processes}.

\subsection{Elastic scattering on electrons}
\label{sec:ScattOnElectrons}
%

We start by analyzing the scattering of neutrinos on electrons, which gets modified in the presence of nonzero neutrino polarizability. The diagram that generates the signal is shown in \cref{fig:Rayleigh:diagram} (left). The end result are two possible electromagnetic (EM) tracks, one due to the emitted photon and the other due to the recoiled electron. 

The differential cross section for $\nu_i(p_\nu) +  e^-(p_e)\to \nu_j(k_\nu) + \g(k_\g) + e^-(k_e)$ is given by
\beq
\label{eq:d:sigma:e}
{\rm d}\sigma_e = \frac{\overline{|{\cal M}_e|}^2}{16(2\pi)^5 E_\nu^{\rm in} m_e} |{\vec k}_e| E_\g \delta(k_\nu^2) {\rm d}E_\g{\rm d}|{\vec k}_e|{\rm d}\Omega_\g{\rm d}\Omega_e\,, 
\eeq
where $\vec{k_e} (E_e)$ is the final state electron three-momentum (energy), $E_\nu^{\rm in}$ the energy of the initial neutrino, $E_\g = |{\vec k}_\g|$ the emitted photon energy, and ${\rm d}\Omega_\g,~{\rm d}\Omega_e$ the differential solid angles for the photon and electron three-momenta, respectively. Out of the six differential variables in \cref{eq:d:sigma:e} we are interested in only three: the photon and electron energies, and the separation angle $\theta_{\g e}$ between $\vec k_\g$ and ${\vec k}_e$. We will use Monte Carlo simulations to numerically integrate the differential cross section in \cref{eq:d:sigma:e} over the other variables, see \cref{sec:Simulations} for details. 

The spin-averaged amplitude squared  in \cref{eq:d:sigma:e} is, in the lab frame, given by
\beq\label{eq:MeSquared}
\overline{|{\cal M}_e|}^2 = \lp\frac{c_\nu^{ij} g_\g \e}{2}\rp^2\frac{Q_\phi^2 E_\g^2 }{ \lp Q_\phi^2 + m_\phi^2 \rp^2} \left( -\frac{|{\vec k}_e|\cos\theta_{e\g}}{m_e} +  \frac{E_e\lp\cos2\theta_{e\g} + 3 \rp}{4m_e} + \frac{m_e\sin^2\theta_{e\g}}{E_e - m_e} \right),
\eeq
where $Q^2_\phi\equiv-q^2_\phi>0$, with $q_\phi = p_\nu - k_\nu$ the four-momentum carried by the virtual pseudoscalar mediator $\phi$.
Note that in the above expression we have traded the dependence of the matrix element on the off-shellness of the virtual photon, $q_\g^2\leq 0$, where  $q_\g = k_e - p_e = q_\phi - k_\g$, for the dependence on the outgoing electron energy $E_e$ and the opening angle $\theta_{e\g}$.  
The cross section is parametrically enhanced whenever $q_\phi^2$ and $q_\g^2$ are small, while $E_\gamma$ is large. Note that the  $q_\g^2\to 0$ limit corresponds to taking $E_e\to m_e$ in  \cref{eq:MeSquared} (i.e., the limit of a very soft virtual photon), leading to the enhanced last term in  \cref{eq:MeSquared}. The $q_\g^2\sim {\mathcal O}(E_e m_e)\ll E_e^2$ limit is achieved for $E_e\gg m_e$, leading to a linear $E_e/m_e\sim |\vec k_e|/m_e$ enhancement  in the first two terms in \cref{eq:MeSquared}.
The pseudoscalar's virtuality $Q_\phi^2+m_\phi^2=-q_\phi^2+m_\phi^2$ is minimized for the smallest value of $Q_\phi^2\sim E_\g (|\vec k_e|\cos\theta_{e\g} - T_e)$, where  $T_e = E_e - m_e$ is the final electron recoil energy. The regime where $Q_\phi^2\simeq0$ occurs when both the outgoing electron and the outgoing photon are soft and may thus be hard to observe. Experimentally perhaps more interesting is the limit of energetic and almost collinear outgoing photon and electron, in which case $Q_\phi^2\sim (m_e E_{\g, e})\ll E_{\g, e}^2$. The challenge in this case could be the small opening angle $\theta_{e\g}$.

First, let us consider the case of a soft electron, $|{\vec k}_e|\sim0$ and $E_e\sim m_e$, so that $q_\g^2\sim0$ and $Q_\phi^2 \sim 2E_\g \sqrt{T_em_e}$, such that the last term in \cref{eq:MeSquared} dominates.
Taking furthermore $Q_\phi^2\gg m_\phi^2$, 
\cref{eq:d:sigma:e} becomes
\beq\label{eq:d:sigma:e:LowRecoil}
{\rm d}\sigma_e \sim \frac{\big( c_\nu^{ij} g_\g \e \big)^2}{ E_\nu^{\rm in} T_e} \frac{E_\g^2\sin^2\theta_{e\g}}{\cos\theta_{e\g}} ~\delta(k_\nu^2)  {\rm d}E_\g{\rm d}T_e{\rm d}\Omega_\g{\rm d}\Omega_e\,,
\eeq
where we neglected constant factors. 
This cross section is parametrically enhanced for small $T_e$ and large separation angles, $\theta_{e\g}$. The experimental challenge is that a very small $E_e$ could be below the detection threshold. Alternatively, if  the neutrino loses only a negligible part of its energy, then both $q_\g^2\sim0$ and $q_\phi^2\sim0$.
In this case, \cref{eq:MeSquared} is proportional to $Q_\phi^2/(m_\phi^4 T_e)$, with $Q_\phi^2\sim{\cal O}(E_\g \sqrt{T_em_e})$. We then obtain ${\rm d}\sigma_e \propto (E_\g/m_\phi)^4$, which goes to 0 as the photon energy goes to zero.

Second, let us consider the limit of an ultra-relativistic outgoing electron, $E_e\gg m_e$, 
and thus $|{\vec k}_e| \simeq E_e$. In this case, the last term in \cref{eq:MeSquared} is negligible, which for $Q_\phi^2\gg m_\phi^2$ gives
\beq
{\rm d}\sigma_e \sim \lp c_\nu^{ij} g_\g \e \rp^2 \frac{E_\g^2 E_e}{E_\nu^{\rm in}m_e^2(1-\cos\theta_{e\g})}\sin^4\lp \frac{\theta_{e\g}}2 \rp ~ \delta(k_\nu^2) {\rm d}E_\g{\rm d}E_e{\rm d}\Omega_\g{\rm d}\Omega_e\,.
\eeq
The above cross section vanishes for $\theta_{e\g}\to0$ and peaks for $\theta_{e\g}\to\pi$, meaning that the two EM tracks in the detector will on average be well separated.  
We quantify these statements in \cref{sec:Signal&Background}, where we also include the detector threshold effects.

\begin{figure}[t]
	\centering
	\includegraphics[width=0.31\linewidth]{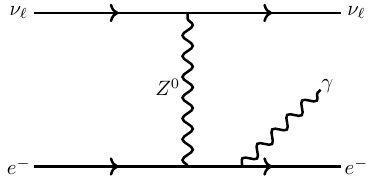}
    \includegraphics[width=0.31\linewidth]{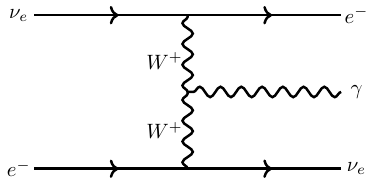}
    \includegraphics[width=0.31\linewidth]{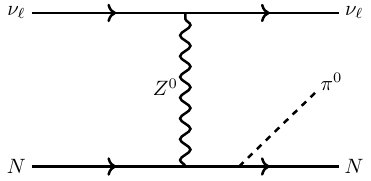}
	\caption{Sample Feynman diagrams for the backgrounds to the searches for enhanced neutrino polarizability: neutral current scattering on electron with a photon bremstrahlung (left); charged current vector fusion (middle); $\pi^0$ emission from argon nucleus (right) --- note that the $\pi^0$ decays to two photons.
	}
	\label{fig:ScattOnEl:Background:diagram}
\end{figure}

%
\subsection{Coherent scattering on argon nucleus}
\label{sec:ScattOnargon}
%

The $\phi$-mediated neutrino scattering on argon nucleus is a sum of coherent and inelastic contributions. We focus on coherent elastic scattering, \cref{fig:Rayleigh:diagram}, which leads to a hard photon signal without hadronic activity, that is easier to distinguish from backgrounds. The differential cross section for $\nu_i(p_\nu) +  \text{Ar}(p_N)\to \nu_j(k_\nu) + \g(k_\g) + \text{Ar}(k_N)$ scattering is 
\beq
\label{eq:dsigma:1EM}
{\rm d}\sigma_N = \frac{1}{(4\pi)^5} \overline{|{\cal M}_N|}^2  \frac{E_\g }{E_\nu} \frac{{\rm d}^3 \vec k_N}{E_\nu^{\rm in} m_{N} E_N} d\Omega_\gamma\,,
\eeq
where $m_N$ is the Ar mass, $E_{\nu}(E_\gamma,E_N)$ is the energy of the outgoing neutrino (photon, Ar nucleus),  $E_{\nu}^{\rm in}$ the energy of the incoming neutrino, and $d\Omega_\g$ the solid angle differential for the outgoing photon.
The spin-averaged amplitude square  is 
\beq
\label{eq:MN2}
\overline{|{\cal M}_N|}^2 = |F(Q_\g^2)|^2 (c_\nu^{ij} g_\g Z \e)^2\frac{Q_\phi^2 E_\g^2}{\lp Q_\phi^2 +m_\phi^2 \rp^2}\frac{\lp 2m_N + T_N \rp }{8T_N}\sin^2\theta_{\rm beam}\,,
\eeq
where $F(Q_\gamma^2)$ is the electromagnetic nuclear form factor, $T_N = E_N - m_N$ is the kinetic energy of the recoiling argon nucleus, and $\theta_{\rm beam}$ the angle between the photon and the neutrino beam directions, in the lab frame. 
We can safely assume that the recoil energy of the nucleus is smaller than its mass, leading to an expression for the cross section as in \cref{eq:d:sigma:e:LowRecoil}, with the replacements $\theta_{e\g}\to\theta_{\rm beam}$, $\vec k_e\to\vec k_N$, $T_e\to T_N$, and including the nuclear form factor:
\beq\label{eq:d:sigma:N:LowRecoil}
{\rm d}\sigma_N \sim (|F(Q_\g)| c_\nu^{ij} g_\g Z \e \big)^2\frac{E_\g^2}{E_\nu^{\rm in} T_N}\frac{\sin^2\theta_{\rm beam}}{\cos\theta_{\rm beam}}~\delta(k_\nu^2)  {\rm d}E_\g{\rm d}T_N{\rm d}\Omega_\g{\rm d}\Omega_N\,,
\eeq
where as in \cref{eq:d:sigma:e:LowRecoil} we also took the $Q_\phi\gg m_\phi$ limit. 

For the electromagnetic form factor we use the Helm form factor~\cite{Duda:2006uk},
\beq\label{eq:HelmFF}
F_{\rm Helm}(Q_\g^2) =  \frac{3j_1(Q_\g R_1)}{Q_\g R_1} e^{-Q_\g^2s^2/2}\,,
\eeq
where $Q_\g^2\equiv-q_\g^2$, $R_1\simeq3.89$\,fm is the effective nuclear radius for argon~\cite{Duda:2006uk}, $s=0.9$ fm is the nuclear skin thickness, and $j_1(x)$ the spherical Bessel function of the first kind.  
In \cref{app:OtherFF} we show the effects of using  other nuclear form factors.

\section{Signal and background processes in the DUNE-ND}
\label{sec:Signal&Background}

Next, let us estimate how well the enhanced neutrino polarizability induced neutrino scatterings on electrons and on argon nuclei, with an emission of a photon,  \cref{eq:processes}, can be distinguished from SM backgrounds. The signal events can be split into two categories, depending on how many EM showers are produced:
\begin{itemize}
    \item 2EM -- two well separated electromagnetic showers are observed, with no associated hadronic activity;
    \item 1EM -- a single electromagnetic shower is observed, with no associated hadronic activity.
\end{itemize}
As we will show below, there are a number of possible sources of backgrounds that also lead to the above two topologies, cf. \cref{fig:ScattOnEl:Background:diagram}, although with distinct kinematics:
neutrino-electron scattering; neutral-current  (NC) events leading to EM showers; or $\nu_e$ charged-current (CC) events.
There are several handles that can be used to mitigate such backgrounds.

In liquid argon time projection chambers, electromagnetic showers initiated by a photon typically display a clear gap from the event vertex.
For simplicity, we will conservatively assume that the electron initiated EM showers cannot be distinguished from the photon initiated ones. 
However, we will also comment 
on the possible improvements, if a separation between the two can be achieved.
The definition of 2EM vs. 1EM category and the specifics of background rejection depend on the capabilities of the DUNE-ND, which we review next.

\begin{figure}[t]
	\centering
	\includegraphics[width=0.8\linewidth]{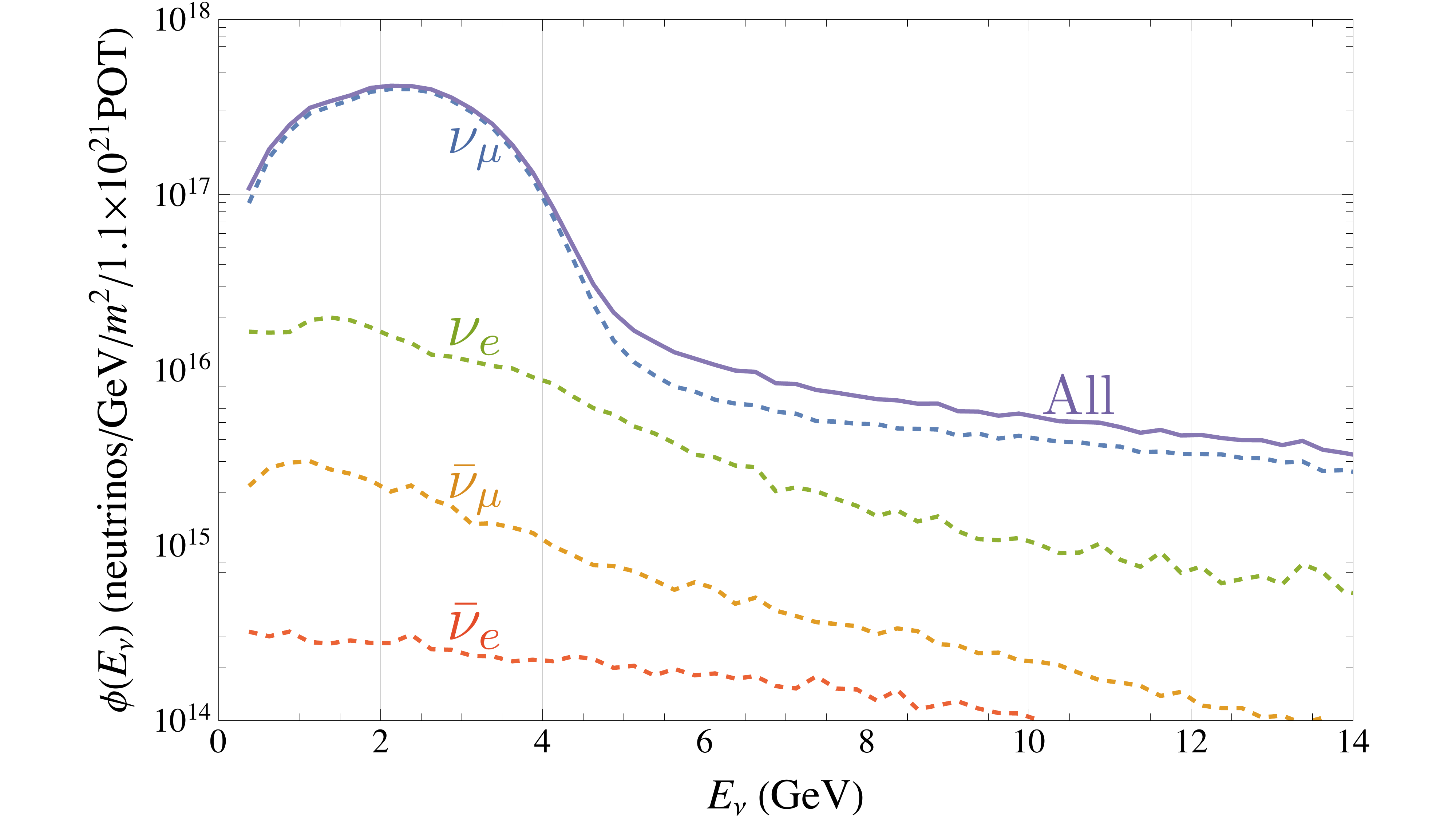}
	\caption{The $\nu_\mu$, $\bar\nu_\mu$, $\nu_e$ and $\bar\nu_e$ fluxes at DUNE-ND~\cite{DUNE:2021cuw} (blue, orange, green and red dashed lines), assuming an exposure of 1 year. The purple solid line indicates the total flux.
	}
	\label{fig:NDflux}
\end{figure}

\subsection{Near Detector}
\label{sec:Detector}
%

The DUNE-ND will be located 574\,m downstream from the neutrino beam production point at Fermilab~\cite{DUNE:2020ypp}. 
The liquid argon detector (ND-LAr) will provide a fiducial volume of 67.2 tonnes of argon, corresponding to a total of $N_{\rm Ar} = 10^{30}$ target argon nuclei. Given the atomic number of argon, $Z = 18$, this yields a total of $N_e = Z N_{\rm Ar} = 1.8 \times 10^{31}$ target electrons.
We do not study the impact of other detectors within the DUNE-ND complex on the search for neutrino polarizability.

The exceptional tracking capabilities of the liquid argon detector will enable the measurement of electromagnetic showers with unprecedented resolution. 
Following the DUNE CDR~\cite{DUNE:2016ymp} and TDR~\cite{DUNE:2020ypp}, we adopt a lower energy threshold of $E_{\rm th} = 30$\,MeV for electrons, muons, and photons, with an assumed energy resolution of 2\%. 
For hadronic tracks, the energy threshold is set to $E_{\rm th} = 100$\,MeV, with an energy resolution of 30\%. 
The directional resolution is taken to be $1^\circ$ for electrons, muons, and photons, and $5^\circ$ for hadrons. 
Finally, the two electromagnetic showers are assumed to be separately reconstructed, if the relative opening angle between the two initial particles, $\theta_{e\gamma}$, is larger than $\theta_{e\g}^\text{th}=3^\circ$. 
We note that our results are not very sensitive to the energy resolution, but rather to the angular reconstruction due to background rejection. In particular, since the threshold  of $\theta_{e\g}^\text{th}=3^\circ$ may well be too optimistic for liquid argon detectors \cite{MicroBooNE:2025ntu,MiniBooNE:2018esg}, we also quantify how the DUNE-ND reach degrades if only larger angular separations will be reconstructable.

The neutrino beam will be produced from proton collisions on a thin graphite target, where PIP-II will allow for  proton energies of up to 120 GeV, with an expected rate of $1.1\times10^{21}$ POT/year for a 1.2 MW beam power.
Using the ancillary files of Ref.~\cite{DUNE:2021cuw}, we plot in \cref{fig:NDflux} the time integrated fluxes $\Phi(E_\nu)$ of muon and electron neutrinos and antineutrinos at the ND as functions of neutrino energy, assuming  1 year integrated exposure and neutrino-enhanced beam running. We use these time integrated fluxes in the numerical study below. 

%
\subsection{Simulation details}
\label{sec:Simulations}
%

To estimate the reach at DUNE we simulate the signal due to $\phi$ mediated neutrino scattering on either electrons or on argon nuclei, \cref{fig:Rayleigh:diagram} (left), as well as the relevant backgrounds, \cref{fig:ScattOnEl:Background:diagram}, for both the 1EM and 2EM signatures. We impose the cuts due to energy thresholds, $E_{\rm th}$, and due to the minimal opening angle, $\theta_{e\g}^\text{th}$, see \cref{sec:Detector}. 

The signal and background rates due to neutrino scattering on electrons and due to 
coherent neutrino scattering on argon
were simulated using {\tt MadGraph v3.5.1}~\cite{Alwall:2014hca}. For this, we implemented a new model containing the SM supplemented by the pseudoscalar $\phi$, \cref{eq:Rayleigh:Lagrangian}, and the argon nucleus modeled as a scalar particle with charge $Z=18$ and mass $m_N=37.52$ GeV. The single-photon electromagnetic vertex of the nucleus was multiplied by the Helm form factor, $F_\text{Helm}(Q_\g^2)$, cf. \cref{eq:HelmFF}.

In the simulation, a scan over the initial neutrino energies $E_{\nu_\ell}\in\big[250\,\text{MeV},10.25\,\text{GeV}]$, where $\nu_\ell=\nu_e, \nu_\mu, \bar \nu_e, \bar \nu_\mu$, was performed in steps of 250 MeV, each time generating $10^5$ events. For signal events the scattering is due to the tree level $\phi$ exchange, \cref{fig:Rayleigh:diagram} (left), where we set $m_\phi = 50$\,MeV in the simulation. For other  $m_\phi$ values the simulated events were reweighted, i.e., each already simulated event was assigned a weight $\big(Q_\phi^2 + m_\phi^2\big)^2/\big( Q_\phi^2 + m_{\phi^\prime}^2 \big)^2$, where $Q_\phi^2=-q_\phi^2$, with $q_\phi$ the momentum flowing through the $\phi$ line in the diagram, and $m_{\phi'}$ the new $\phi$ mass. To avoid infrared singularities, we imposed  in the simulation a weak cut on the photon transverse momentum of $p_{T,\g}\gtrsim1$\,MeV. We checked that this cut is loose enough such that it does not affect the values for the cross sections,  after the detector thresholds are applied. 

In principle, the tree level $\phi$ exchange induced non-coherent neutrino-argon scattering could also be a source of signal events. 
However, in this case the presence of visible hadronic activity makes such signal events very similar to the copious SM interaction induced neutrino-argon events.
Due to the  low expected signal-to-background ratio, we do not include non-coherent neutrino-argon scattering in our estimates for the signal events. 

For the calculation of background rates due to 
quasi-elastic and inelastic neutrino-argon scattering we used  {\tt NuWro}~\cite{Golan:2012rfa}. We generated $3\times10^6$ events for both $\nu_e$-CC  and total $\nu$-NC samples.
These samples include quasi-elastic, resonant, meson-exchange current, and deep inelastic scatterings, as well as coherent pion production. 
Neutral pions in the final state were decayed to two photons. 
At the end of the simulation we imposed the energy and angular separation thresholds from  \cref{sec:Detector}, with electrons and photons treated as indistinguishable objects generating the EM showers.  
 
In  principle, neutrino tridents $\nu {\rm Ar}\to e^+ e^- \nu {\rm Ar}$ could also present a background to both 2EM and 1EM signal topologies. 
However, the cross sections for neutrino tridents are $\mathcal{O}(10^4)$ times smaller than for the typical SM neutrino-argon CC and NC scatterings.
Furthermore,  electron-photon discrimination, if included in the analysis pipeline, should be quite effective in mitigating this background.
We therefore expect the neutrino tridents not to be a significant source of background events, and thus neglect them in our estimates.

The differential distribution for different types of events are obtained by convolving the corresponding cross sections with the time integrated neutrino fluxes, 
\beq\label{eq:EventDistribution:general}
\frac{{\rm d}N}{{\rm d}X} = N_T 
\times \sum_\nu \int {\rm d}E_\nu\, \phi_\nu(E_\nu)\, \frac{{\rm d}\sigma_\nu}{{\rm d}X}(E_\nu)\,,
\eeq
where the sum is over the four relevant neutrino species $\nu_e, \nu_\mu, \bar \nu_e$, and $\bar\nu_\mu$. The time integrated fluxes $\phi_\nu(E_\nu)$ are shown in \cref{fig:NDflux}, and $N_T$ is the total number of target particles in the detector (see \cref{sec:Detector}). 
The differential variable ${\rm d}X$ can be one-dimensional or two-dimensional, for instance $\text{d}\theta_{e\gamma}$ or $\text{d}\theta_{e\gamma} \text{d}E_{\rm EM}$, where $E_{\rm EM}$ is the total electromagnetic energy deposited in the neutrino scattering event. The cross sections $\sigma_\nu$ for the two types of events, 2EM and 1EM, are summed over the signal and the background contributions. 
We discuss next how one can experimentally distinguish between signal and background, for each of the two types of events.

\begin{figure}[t]
	\centering
    \includegraphics[width=0.9\linewidth]{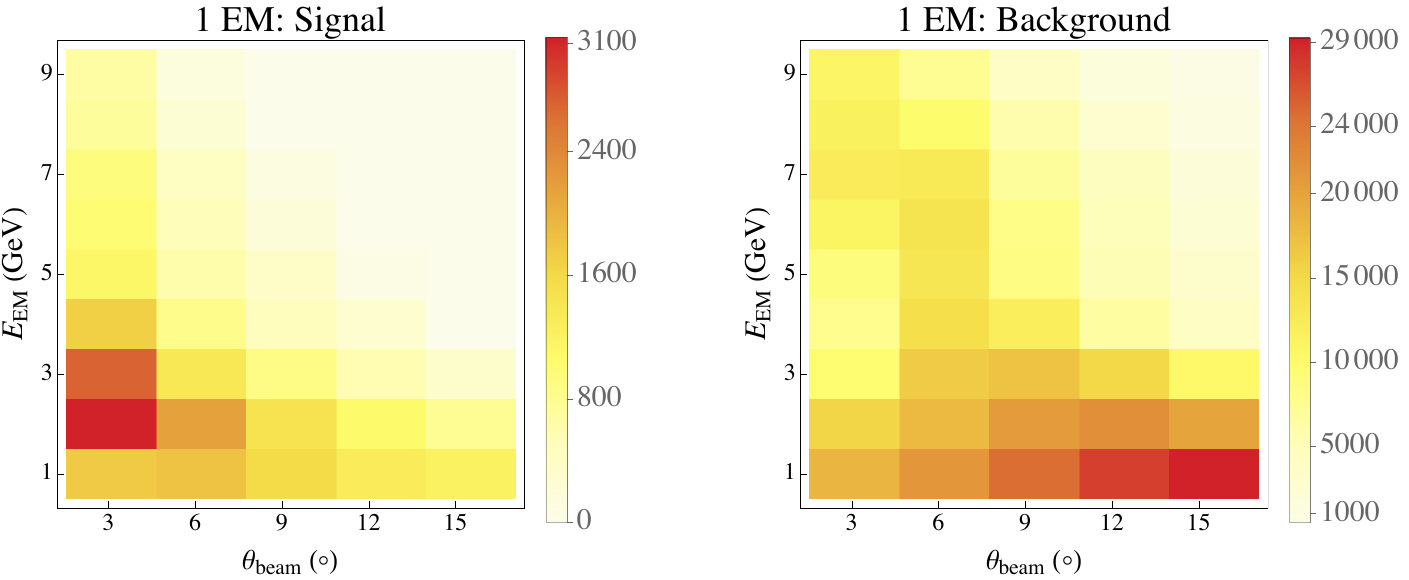}
	\caption{Kinematical distributions for the 1EM shower topology of expected number of signal (left) and background (right) events after 1 year exposure, as functions of EM shower energy $E_{\rm EM}$ and the  angle of EM shower with respect to the beam $\theta_\text{beam}$. The bins are labeled by its lower boundary, e.g., the $\theta_\text{beam}=3\degree$ bin corresponds to the range $\theta_\text{beam}\in (3\degree, 6\degree]$, and similarly for the energy, e.g., the $E_\text{EM}=1\,$GeV bin spans the interval $E_\text{EM}\in (1,2]\,$GeV.
    The signal spectrum was computed for the benchmark value of neutrino polarizability $\tilde\alpha_\nu = 10^{-6}/(4 m_\phi^2)$, setting the mediator mass to $m_\phi= 50$ MeV, see text for details. The legends show the number of expected events in each bin.
    }
	\label{fig:SvsB:1EM}
\end{figure}

\subsection{One electromagnetic shower: signal and backgrounds}
\label{sec:Signal&Bkg:1EM}
 
The signal 1EM events are mainly due to coherent neutrino scattering on argon nuclei, which leads to the emission of an energetic photon, see \cref{fig:Rayleigh:diagram}. The photon is expected to be emitted in the very forward direction, which distinguishes it from dominant backgrounds. As observables, we focus on the energy deposited in the EM shower, $E_{\rm EM}$, and the angle of the EM shower relative to the beam axis, $\theta_\text{beam}$. 
In \cref{fig:SvsB:1EM}, we show the 2D distribution of the signal (left panel) and background events (right panel) for a representative benchmark point, 
\beq\label{eq:coupling:benchmark}
\tilde{\alpha}_\nu=10^{-4} \mbox{ GeV}^{-1}, \quad m_\phi=50 \mbox{ MeV}\,.
\eeq
The benchmark value of  $\tilde{\alpha}_\nu$ is comparable to the DUNE-ND reach in our most conservative estimates, see \cref{sec:reach}. In terms of coupling to neutrinos and photons, the above benchmark corresponds to $c_\nu g_\g = 10^{-6}\,{\rm GeV}^{-1}$, see Eq. \eqref{eq:tilde:alpha:nu}.

There are a few sources of backgrounds for the 1EM topology. 
First, $\nu_e$-CC scattering results in a final state electron, and thus a single EM shower.
Additionally, $\nu$-NC events may yield photons that are sometimes classified as single electromagnetic showers.
Examples include a $\pi^0$ decaying into collinear photons, or $\pi^0\to 2\gamma$ decays in which one of the photons is below reconstruction threshold.
While the rate for these backgrounds are very large, requiring the shower to be in the very forward direction and imposing no observable hadronic activity eliminates the vast majority of the events.

Another contribution to the background arises from neutrino-electron scatterings, $\nu_\ell e^-\to\nu_\ell e^-$, which also results in a single EM shower. The cross section for this process is smaller than the one for coherent scattering on argon, however, the final state kinematics is very similar to our signal events. In particular, the angular distribution of the EM showers peaks in the very forward region, and there is no visible hadronic activity. \Cref{fig:SvsB:1EM} (right) shows the sum of all backgrounds for the 1EM event types, after imposing all the kinematics and hadronic activity cuts. We see that the kinematics of signal and background events are quite distinct, with still about an order of magnitude more background than signal events expected in the regions of the $(\theta_\text{beam},E_\text{EM})$ plane with the larger concentration of signal events, for the benchmark value of the neutrino polarizability.

\begin{figure}[t]
	\centering
    \includegraphics[width=0.9\linewidth]{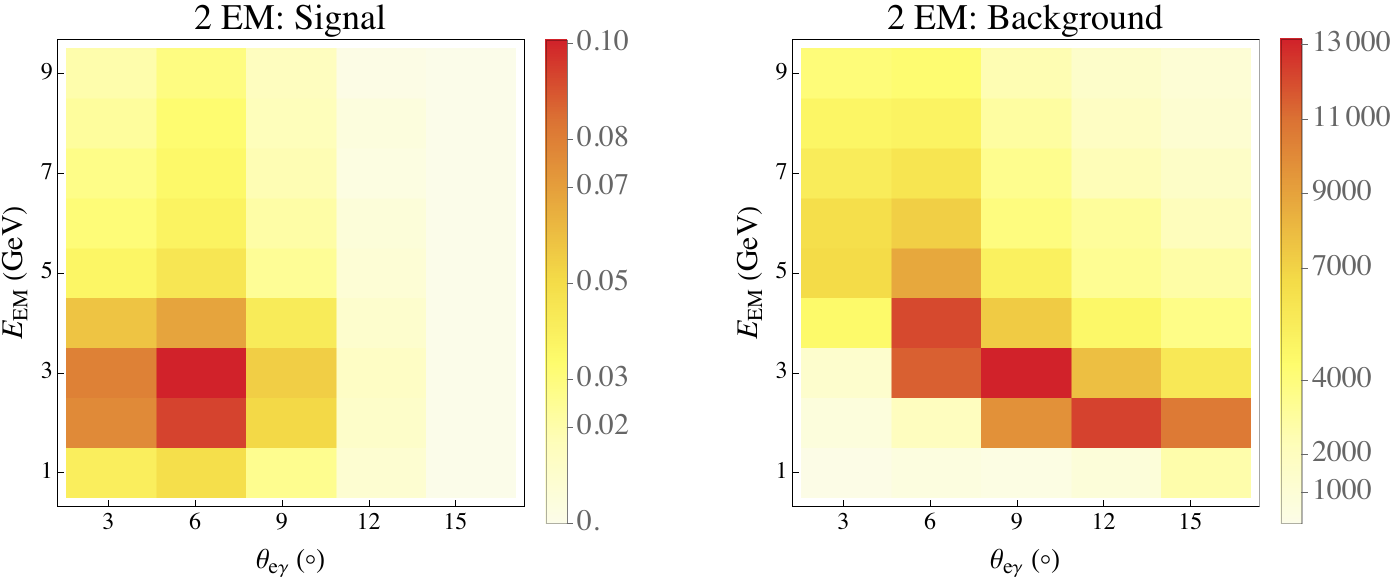}
	\caption{Same as \cref{fig:SvsB:1EM}, for the 2EM event types.
    }
	\label{fig:SvsB:2EM}
\end{figure}

\subsection{Two electromagnetic showers: signal and backgrounds}
\label{sec:Signal&Bkg:2EM}

The minimal experimental requirements for the 2EM event types is that both showers pass the lower bound energy cut, and that the separation angle is larger than the detector resolution. 
The effect of these cuts can be captured by focusing on the following two observables: the total energy deposited via EM showers, $E_{\rm EM} \equiv E_\g + E_e$, and the separation angle between the electron and the photon, $\theta_{e\g}$. 
In \cref{fig:SvsB:2EM} we show the 2D distributions of the signal (left) and background (right) events in the $(\theta_{e\g},E_{\rm EM})$ plane. 
The main feature of the signal is a peaking distribution at $E_{\rm EM}\sim 3$ GeV, while the separation angle of the two showers extends as far as $\theta_{e\g}\sim10\degree$. The background has a similar peaking structure in $E_{\rm EM}$, but in general with a larger separation angle $\theta_{e\gamma}$.

The SM background processes for the 2EM signature can come both from scattering on electron and scattering on argon nuclei. 
In the former case, this can be achieved by either emitting a photon by bremsstrahlung from the outgoing electron leg, or from an internal gauge boson line, as in the diagrams shown in \cref{fig:ScattOnEl:Background:diagram}. 
The processes involving weak charged-current (CC) interactions can only be initiated by an electron neutrino scattering on an electron, while neutral-current (NC) scatterings can be initiated by any of the four neutrino types.

For neutrino-argon scattering backgrounds, the 2EM signature comes from neutrino-nucleus interactions with no observable hadronic activity.  
This is typically the case for CC events that produce one electron and one $\pi^0$, where one of the $\pi^0$ photons is either below threshold or cannot be isolated due to small opening angle with respect to other EM showers. 
The same type of backgrounds can be obtained in the case of a muon produced below detection threshold.
Neutral-current interactions between neutrinos and nucleus can also lead to double EM showers, if $\pi^0$ is produced in the scattering process, where $\pi^0$ then decays to $2 \g$, typically with a large opening angle.

%
\section{DUNE's reach}
\label{sec:reach}
%

\begin{figure}[t]
	\centering
	\includegraphics[width=0.8\linewidth]{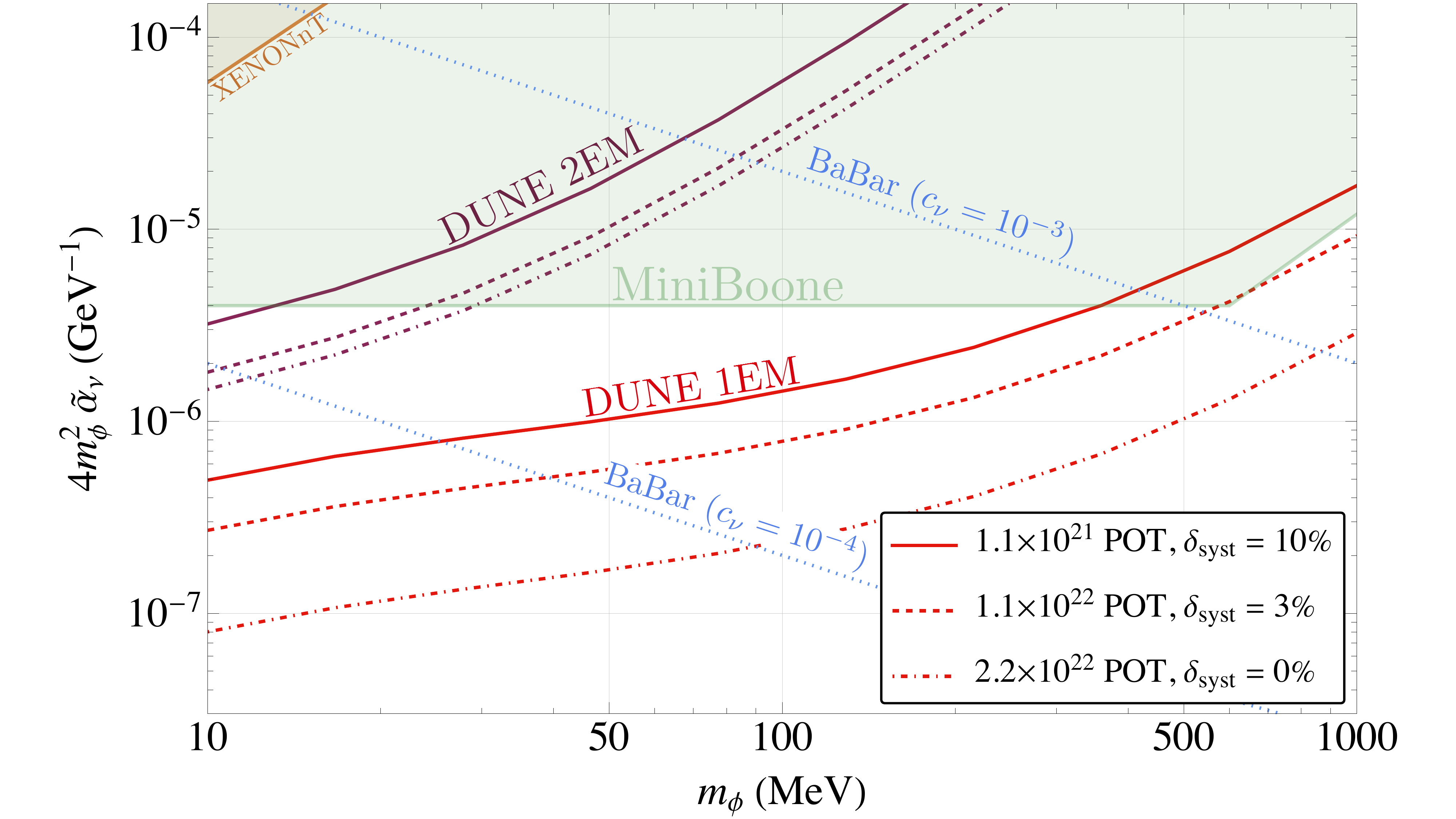}
	\caption{Bounds on rescaled neutrino polarizability as function of $m_\phi$. The red (dark red) lines show the reach of DUNE-ND using 1EM (2EM) event types, with solid (dashed, dotted) lines corresponding to Case I (Case II, Stat. only) scenarios for the values of statistics and systematic uncertainties, \cref{eq.CaseI,eq.CaseII,eq.Stat:only}.  The shaded regions indicate the present bounds from MiniBooNE (light green) and XENONnT (orange). The dotted blue lines show the BaBar constraint for two benchmark values of $c_\nu$, see text for details. 
    }
	\label{fig:BoundsVSmassphi}
\end{figure}

In order to estimate the DUNE-ND reach, we construct  
$\chi^2$ functions for each bin in the $(\theta_{e\gamma},E_{\rm EM})$ plane for 2EM, and in the $(\theta_{\rm beam},E_{\rm EM})$ plane for the 1EM signature, respectively. In both cases the 2D  bins have a size $(3\degree,1~{\rm GeV})$. For the construction of $\chi^2$ values we assume the Asimov set, i.e., that in each bin the observed 
number of events match the expected SM background values. A nonzero polarizability signal, giving $N_S^i(\tilde\alpha_\nu, m_\phi)$ expected events in the $i-$th bin would thus give the following contribution to the $\chi^2$ function,
\beq\label{eq:ChiSquared}
\chi_i^2(\tilde\alpha_\nu, m_\phi) = \lp \frac{N_S^{i}(\tilde\alpha_\nu, m_\phi)}{\sigma_{B,{\rm tot}}^{i}} \rp^2 \,,
\eeq
where we assumed that the Gaussian limit can be used. The uncertainty on the expected number of events in each bin has statistical and systematic components that we sum quadratically, 
\beq
\sigma_{B,{\rm tot}}^{i} = \sqrt{(\sigma_{B,{\rm stat}}^{i})^2 + (\sigma_{B,{\rm syst}}^{i})^2}.
\eeq
The statistical error is $\sigma_{B,{\rm stat}}^{i} = \sqrt{N_B^{i}}$, with $N_B^{i}$ the expected number of SM events in $i$-th bin, while for the systematic uncertainty  $\sigma_{B,{\rm syst}}^{i} = \delta_{\rm syst} N_B^{i}$. The total $\chi^2$ is then obtained by summing over all the bins, $\chi^2=\sum_i \chi_i^2$, for both 1EM and 2EM signatures.  We consider three representative cases for the collected statistics (number of protons on target) and the systematic errors,
\begin{align}
\label{eq.CaseI}
\text{Case I:}& \quad 1.1\times 10^{21}\,\text{POT}, \quad \delta_\text{syst}=10\%,
\\
\label{eq.CaseII}
\text{Case II:}& \quad 1.1\times 10^{22}\,\text{POT}, \quad \delta_\text{syst}=3\%,
\\
\label{eq.Stat:only}
\text{Statistics only:}& \quad 2.2\times 10^{22}\,\text{POT}, \quad \delta_\text{syst}=0\%.
\end{align}
Case I corresponds to roughly 1 year of data taking in the nominal DUNE-ND run, with a rather conservative estimate for the size of systematic errors. 
Case II, on the other hand, corresponds to 10$\times$ larger statistics, and an aggressive benchmark for the size of the systematic uncertainties. 
The achievable DUNE-ND reach most likely lies in between these two scenarios. As a comparison, we also show the ultimate reach with the expected DUNE-ND statistics, if the systematic uncertainties were negligible (``Statistics only'' dot-dashed lines in \cref{fig:BoundsVSmassphi}).
This latter case is only for illustration purposes, showing that in all realistic scenarios the search is expected to be systematics dominated. 

The highest sensitivity to neutrino polarizability signal comes from similar deposited electromagnetic energy, $E_{\rm EM}\sim3$\,GeV, for both 1EM and 2EM signatures (see \cref{fig:SvsB:1EM,fig:SvsB:2EM}, as well as the additional 2D distributions shown in \cref{sec:app:S:to:B}). For 2EM the highest sensitivity to the signal is for small angular separations between the two showers, $\theta_{e\gamma}\sim 3^\circ$, i.e., close to the assumed resolution. Similarly, for 1EM signature the highest sensitivity is for small $\theta_{\rm beam}$, i.e., for forward EM showers. 

The expected 90\%\,CL bounds on neutrino polarizability follow from requiring $\chi^2(\tilde\alpha_\nu) \leq 2.7$, which for  $m_\phi = 50$\,MeV gives, 
\begin{align}
\label{eq:2EM:project}
    {\rm 2EM}: &\qquad\tilde\alpha_\nu \leq (18,~~10,~~~8.3)\times10^{-4}~{\rm GeV}^{-3}\,, 
    \\
    \label{eq:1EM:project}
    {\rm 1EM}:& \qquad \tilde\alpha_\nu \leq (1.0,~0.56,~0.17)\times10^{-4}~{\rm GeV}^{-3}\,,
\end{align}
for Case I, Case II and Statistics only, respectively, see \cref{eq.CaseI,eq.Stat:only}. In all three cases the 1EM search is significantly more constraining, leading to at least an order of magnitude more stringent constraints on $\tilde \alpha_\nu$. In terms of $\phi$ couplings to the neutrinos and photons, \cref{eq:tilde:alpha:nu}, the above bounds are 
\begin{align}
    {\rm 2EM}:& \quad c_\nu g_\gamma=4m_\phi^2\times\tilde\alpha_\nu \leq (18,~~10,~~~8.3)\times10^{-6}~{\rm GeV}^{-1}\,, \\
    \label{eq:1EM:cnucgamma}
    {\rm 1EM}: &\quad c_\nu g_\gamma= 4m_\phi^2\times\tilde\alpha_\nu \leq (1.0,~0.56,~0.17)\times10^{-6}~{\rm GeV}^{-1}\,,
\end{align}
in all case taking $m_\phi=50$\,MeV.  

The above results indicate that the main sensitivity to neutrino polarizability is expected to be  from the 1EM signature. Furthermore, the sensitivity of the 2EM signature quickly degrades, if only larger separation angles between the two electromagnetic showers are observable. For instance, while the projections in \cref{eq:2EM:project} assume a threshold separation angle of $\theta_{e\gamma}^\text{th}=3\degree$, which gives $\tilde\alpha_\nu \leq 18\times10^{-4}~{\rm GeV}^{-3}$ for the Case I scenario, thresholds of $\theta_{e\gamma}^\text{th}=6\degree(9\degree,12\degree)$ would lead to $\tilde\alpha_\nu \leq 32(68,670)\times10^{-4}~{\rm GeV}^{-3}$. For comparison, the separation thresholds between two photon-induced showers have been estimated to be $13\degree$ at MiniBooNE \cite{MiniBooNE:2018esg} and $20\degree$ at MicroBooNE \cite{MicroBooNE:2025ntu}. Recent studies show the possibility to reach as low as $\theta_{e\gamma}^\text{th}=5\degree$ at MicroBooNE~\cite{MicroBooNE:sepangle}, indicating that a similar or better performance of the DUNE-ND would render the 2EM search phenomenologically relevant. 
In contrast, the projected sensitivity reach from the 1EM signature is less sensitive to the exact value of the $\theta_\text{beam}^\text{th}$ threshold. Increasing the threshold value from  $\theta_\text{beam}^\text{th}=3\degree$, which was used in \cref{eq:1EM:project},  to the larger values $\theta_\text{beam}^\text{th}=6\degree (9\degree)$, reduces the projected sensitivity from $\tilde\alpha_\nu \leq 1.0\times10^{-4}~{\rm GeV}^{-3}$ to  $\tilde\alpha_\nu \leq  2.2(3.8)\times10^{-4}~{\rm GeV}^{-3}$, for the Case I scenario. Naively, one also expects that smaller thresholds may be achieved for the $\theta_\text{beam}$ angle than for $\theta_{e\gamma}$: a single EM cluster averaging out into some unique direction is a simpler problem than a separation between two overlapping EM showers. While our use of a small $\theta_\text{beam}^\text{th}$ threshold in the projections may well be warranted, our results should be viewed with this caveat in mind, until realistic estimates of the $\theta_\text{beam}^\text{th}$ threshold based on detailed simulations of the DUNE-ND detector performance become available. 

For other $\phi$ masses, not just for $m_\phi=50$\,MeV as in Eqs. \eqref{eq:2EM:project}-\eqref{eq:1EM:cnucgamma},  the expected DUNE-ND reach is shown   in \cref{fig:BoundsVSmassphi} with purple and red lines, respectively. 
For these, we used the rescaling prescription described in \cref{sec:Simulations} to compute the bounds for values of masses spanning from 10 MeV to 1 GeV. Note that for significantly lighter masses, $m_\phi\lesssim1$ MeV, there are stringent constraints from cosmological and astrophysical observables~\cite{Bansal:2022zpi}, excluding any parameter space that could be explored by the DUNE-ND for light $m_\phi$.

In contrast, in the 10 MeV to 1 GeV mass range for $m_\phi$ there is significant parameter space that can be probed by the DUNE-ND for the first time. In \cref{fig:BoundsVSmassphi}, the shaded regions denote present exclusions from other relevant terrestrial experiments, MiniBooNE and XENONnT~\cite{XENON:2022ltv, Bansal:2022zpi}. 
MiniBooNE observes the scattering of neutrinos of ${\cal O}(1)$ GeV energy on CH$_2$ nuclei. 
The $\phi$ exchange induced polarizability interaction is thus described well by the EFT operator already for  $m_\phi\lesssim1$ GeV. In the EFT regime, MiniBooNE sets a bound on $\tilde \alpha_\nu$, explaining the reduction of bound on $c_\nu g_\gamma=4m_\phi^2 \tilde \alpha_\nu$ in \cref{fig:BoundsVSmassphi} for heavier masses. For smaller values of $m_\phi$, on the other hand, the bound on $c_\nu g_\gamma$ becomes independent of $\phi$ mass, as expected, and also confirmed by the results shown in \cref{fig:BoundsVSmassphi}. 
In contrast, XENONnT places a constraint directly on the neutrino polarizability, $\tilde \alpha_\nu$, by comparing the observed solar neutrino scattering  signal rates with the SM expectations. 
For the full range of $\phi$ masses considered in the XENONnT analysis, the EFT description is valid and thus the bound on $\tilde \alpha_\nu$ does not depend on $m_\phi$.

In \cref{fig:BoundsVSmassphi}, we also show the BaBar bounds as dotted lines~\cite{BaBar:2017tiz}, with regions above the lines excluded. The polarizability signal at BaBar (or other $e^+e^-$ colliders) would be from associated $\phi$ production, $e^+e^-\to\g\phi,~\phi\to\g\g$. If the $\phi\to \g\g$ channel is the dominant decay mode, so that  $\text{Br}(\phi \to \g\g)\simeq 100\%$, then the signal only depends on the value of $c_\g$. For a given mass $m_\phi$, the BaBar analysis then translates into a bound on $\tilde \alpha_\nu$, if a particular value of couplings to neutrinos, $c_\nu$, is assumed. In \cref{fig:BoundsVSmassphi}, we show the exclusions for two representative values, $c_\nu = 10^{-3}$ and $c_\nu = 10^{-4}$. We see that for smaller values of $c_\nu$ the constraint on $\tilde \alpha_\nu$ becomes correspondingly weaker. For larger values of neutrino couplings there is an additional effect that further weakens the bounds, specifically that the branching ratio for decays to neutrinos becomes non-negligible. For instance, for $c_\nu\gtrsim10^{-4}$, the branching ratio of $\phi\to\nu\nu$ is already larger than $\sim1\%$. 

\begin{figure}[t]
	\centering
	\includegraphics[width=0.48\linewidth]{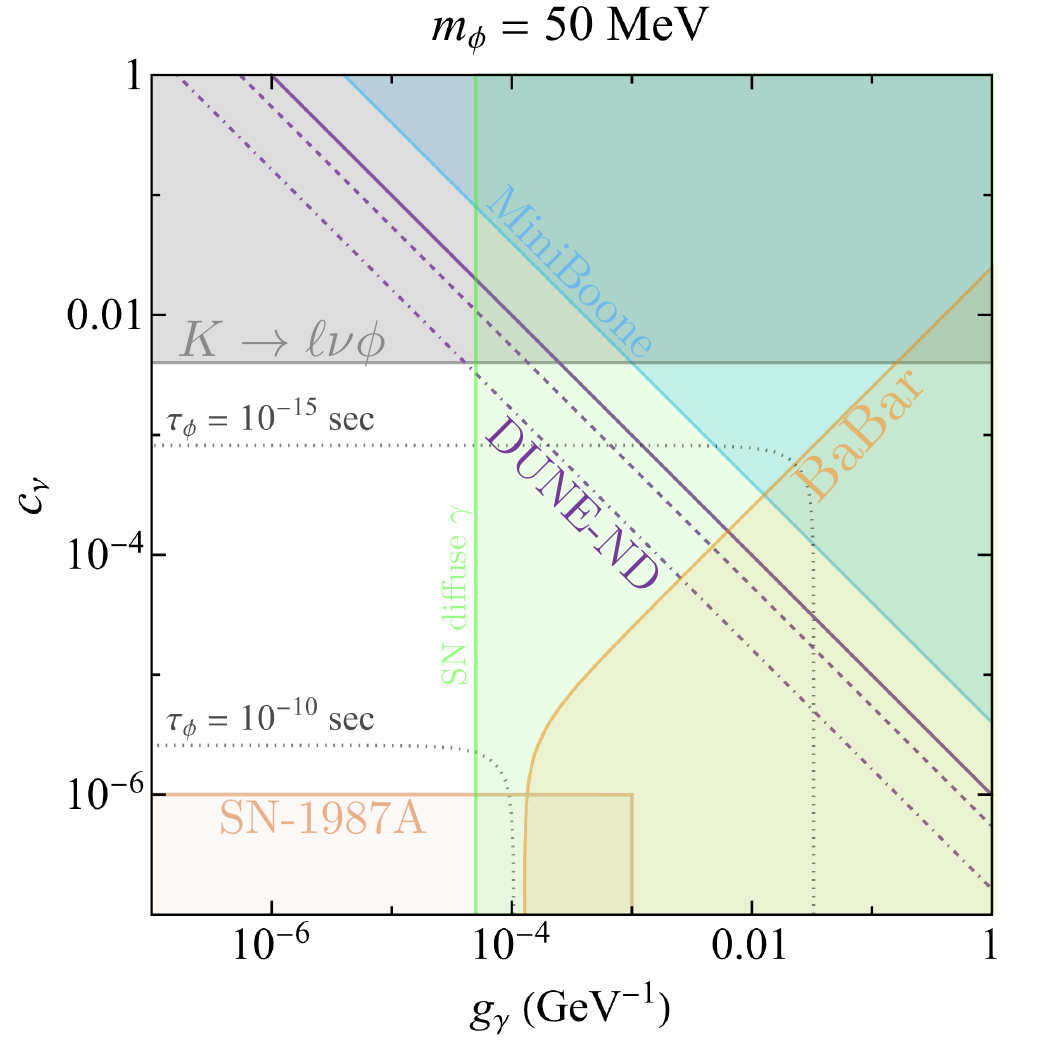}
    \includegraphics[width=0.48\linewidth]{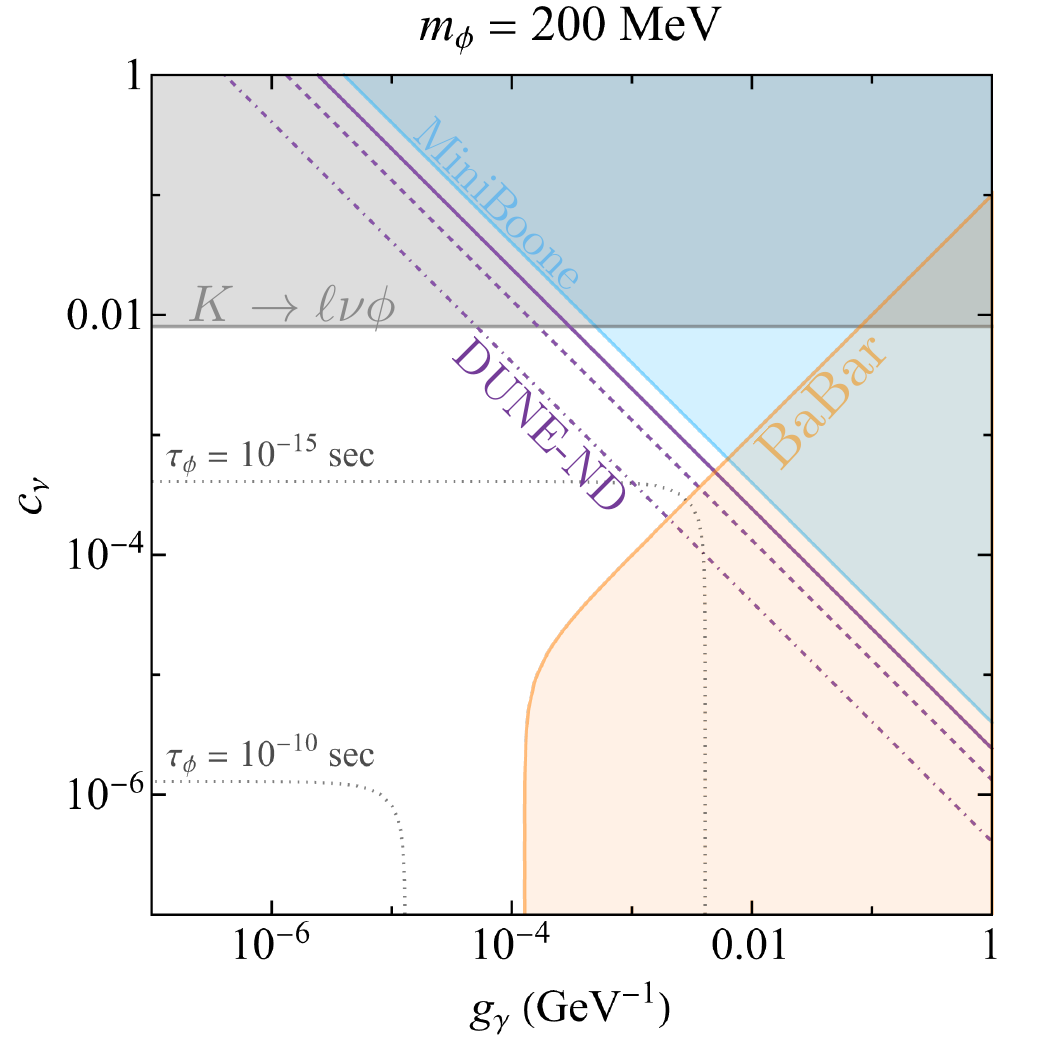}

    ~

    \includegraphics[width=0.48\linewidth]{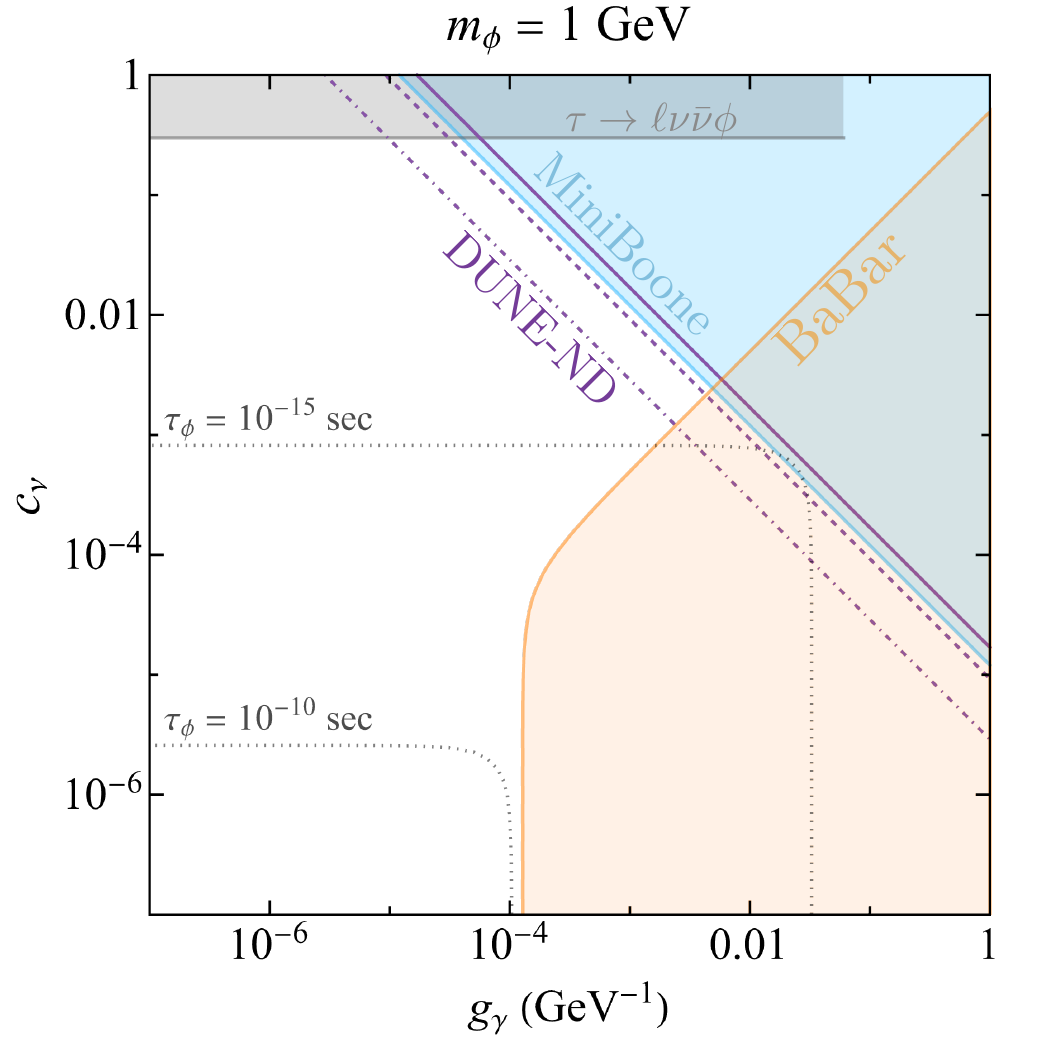}
	\caption{Bounds on the neutrino and photon couplings, $c_\nu$ and $g_{\phi\g}$, for three benchmarks of the $\phi$ mass, $m_\phi = 50, 200, 10^3$ MeV. The purple lines indicate the result of this work for the three representative cases, considering the strongest bound coming from the 1EM signature. }
	\label{fig:BoundsAtBenchmarks}
\end{figure}

A different representation of the DUNE-ND reach compared to current constraints is shown in \cref{fig:BoundsAtBenchmarks}, where we fix $m_\phi$ to three representative values, $m_\phi = 50,\,200,\,10^3$ MeV, and plot the excluded regions in the plane defined by the photon coupling, $g_{\g}$, and the flavor-universal neutrino coupling, $c_\nu$. 
The three panels in \cref{fig:BoundsAtBenchmarks} highlight a number of experimental constraints that probe just the $c_\nu$ or the $g_\g$ coupling. Specifically, the bounds from semileptonic kaon decays~\cite{Blinov:2019gcj}, $\tau$ lepton lifetime~\cite{Lessa:2007up,Pasquini:2015fjv} and the SN cooling rate~\cite{Lucente_2020,Caputo:2022rca} set bounds on $c_\nu$, while the limit from the SN diffuse $\g$-ray spectrum~\cite{Caputo:2021rux} constrains $g_\g$ (in the ranges of the two couplings of interest to us).

In all three panels in \cref{fig:BoundsAtBenchmarks}, the purple lines denotes the projected reach of DUNE-ND, using the 1EM signature with the three benchmarks described in \cref{eq.CaseI,eq.CaseII,eq.Stat:only}. 
For light masses, $m_\phi\lesssim50$ MeV, a large part of the relevant parameter space is already excluded by the astrophysical constraints. The $\phi$ production in astrophysical systems become kinematically inaccessible for masses larger than ${\mathcal O}(100)$\,MeV, as even the SN environment cannot reach high enough temperatures. For $m_\phi\gtrsim200$\,MeV, therefore. the probe of choice for enhanced neutrino polarizability models are the terrestrial experiments, with DUNE-ND projected to go beyond the current constrains. 

%
\section{Conclusions}
\label{sec:Conclusions}
In this paper, we performed a sensitivity study of DUNE's Near Detector to neutrino polarizability. We focused on models of enhanced neutrino polarizability where a light mediator $\phi$ couples to both neutrinos and photons. Neutrino scatterings on electrons and nuclei via tree level $\phi$ exchanges result in one electromagnetic (1EM) and two electromagnetic shower (2EM) signatures, respectively, see \cref{fig:Rayleigh:diagram} (left). For both signatures, the kinematics of the signal events are distinct from the dominant backgrounds. Taking experimental thresholds into account, the 2EM signal peaks more toward small separation angles between electrons and photons and toward small deposited total electromagnetic energy, than the main backgrounds do. Similarly, the 1EM signal events peak more toward small opening angles between the outgoing electron and the incoming beam direction. For the same value of neutrino polarizability (and the mediator mass), the signal-to-background ratio is higher for the 1EM, so we expect this to be the more sensitive search strategy. The expected reach at DUNE-ND is shown in \cref{fig:BoundsVSmassphi} and \cref{fig:BoundsAtBenchmarks}, and would probe new regions of parameter space.

While the main goal of the presented phenomenological analysis was to determine whether DUNE-ND can be sensitive to the still-allowed parameter space in models of enhanced neutrino polarizability, the analysis made several assumptions that could be improved in the actual experimental search. Perhaps the most important is to include the discriminating power of the liquid argon detector between electron- and photon-initiated electromagnetic showers. This could significantly improve the sensitivity of 2EM-type events to neutrino polarizability. 

Another very important aspect is how well the systematic uncertainties on the SM background can be controlled. Here, the most crucial element is the shape information in the 2D distributions. Finally, the binning used in our analysis could be optimized to enhance sensitivity to the signal, or an unbinned maximum likelihood estimator could be employed. In short, there may be significant improvements possible, and a more detailed analysis of the experimental reach by the DUNE Collaboration seems warranted.  

{\bf Notes added:} While this work has been reviewed and approved as a DUNE theory paper by the DUNE Collaboration, the results presented herein represent the views of the authors and not of the DUNE Collaboration as a whole. 

During the final stages of this paper's completion, two other papers appeared that consider similar, but not identical, topics. 
Ref. \cite{Gehrlein:2025tko} focuses on monophoton signatures from neutrino scattering on nuclei arising from neutrino polarizability. 
Ref. \cite{Dutta:2025fgz} considers monophoton signatures 
from neutrino and/or dark matter scattering on nucleus/nucleon via the exchange of a $Z^\prime$ and a light scalar.  


%
\section*{Acknowledgments}
We are grateful to Justin Evans and Justo Mart\'{i}n-Albo for helpful discussions.
S.C., G.P. acknowledge support by the DOE grant DE-SC0007983. J.Z. and A.S. acknowledge support in part by the DOE grant de-sc0011784, and NSF grants OAC-2103889, OAC-2411215, and OAC-2417682. M.T. acknowledges support by Next Generation EU, as part of Piano Nazionale di Ripresa e Resilienza (PNRR), Missione 4, Componente 2, Investimento 1.2 - CUP I13C25000150006. A.A.P. was supported in part by the US Department of Energy grant DE-SC0024357. This paper has been authored by Fermi Forward Discovery Group, LLC under Contract No. 89243024CSC000002 with the U.S. Department of Energy, Office of Science, Office of High Energy Physics. 
%

%
%

\begin{appendix}

\section{Other form factors}
\label{app:OtherFF}
In this appendix, we quantify the impact of using different nuclear form factor models instead of the Helm form factor in \cref{eq:MN2}.

\begin{figure}[t]
	\centering
	\includegraphics[width=0.7\linewidth]{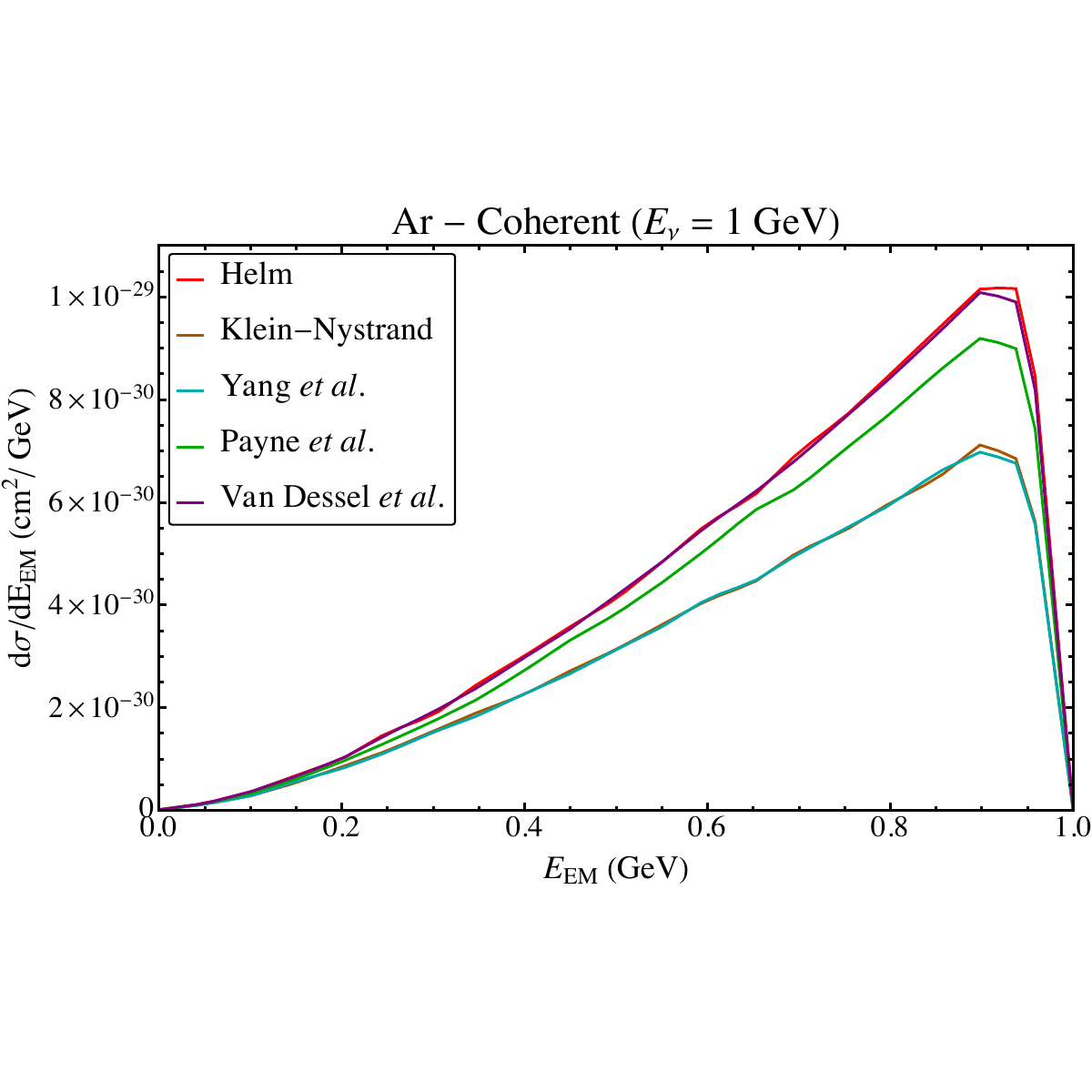}
	\vspace{-2cm}\caption{Differential cross section in the emitted photon energy, for a neutrino with $E_\nu = 1$ GeV scattering coherently on argon, using the Helm \cite{Helm:1956zz}, Klein-Nystrand \cite{Klein:1999}, Yang \textit{et al.} \cite{Yang:2019pbx}, Van Dessel \textit{et al.} \cite{VanDessel:2020epd}, and Payne \textit{et al.} \cite{Payne:2019wvy} form factors.
	}
	\label{fig:FF_Cmprsn}
\end{figure}

In \cref{fig:FF_Cmprsn}, we show a comparison of the differential cross section in the emitted photon energy, for an incoming neutrino energy of 1 GeV scattering coherently off an argon nucleus, using the Helm, Klein-Nystrand \cite{Klein:1999}, RMF \cite{Yang:2019pbx}, HF-SkE2 \cite{VanDessel:2020epd} and the coupled-cluster approach \cite{Payne:2019wvy} form factors. The Helm form factor (used as default in \cref{eq:HelmFF}) and the Klein-Nystrand form factor are the two most widely used phenomenological form factors in CE$\nu$NS analyses. For the Klein-Nystrand form factor, the nuclear density is modeled as a Woods-Saxon distribution with radius $R_A$ and convoluted with a Yukawa potential of range $a_k$, giving \cite{Klein:1999} 
\beq\label{eq:KNFF}
F_{\rm KN}(Q^2) = \lp \frac{3j_1(QR_A)}{QR_A} \rp \lp \frac{1}{1+a_k^2\,Q^2} \rp \,.
\eeq

The remaining nuclear form factors are based on nuclear structure calculations. The HF-SkE2 form factor \cite{VanDessel:2020epd} is numerically calculated by solving the Hartree–Fock equations with a Skyrme potential for the nucleus. The prediction for the form factor in Ref.~\cite{Payne:2019wvy} was obtained using coupled-cluster theory. The RMF model \cite{Yang:2019pbx} employs the relativistic mean-field approach, incorporating the properties of finite nuclei and neutron stars. 

The choice of form factor introduces a systematic uncertainty in the predicted signal rate. As shown in \cref{fig:FF_Cmprsn}, the spread among different models results in an overall normalization difference of about 10--20\%. This is consistent with the expected size of such systematics in our signal analysis.

\section{Signal-to-background estimates}
\label{sec:app:S:to:B}
In this appendix, we collect additional figures that highlight the differences between the kinematical distributions of the signal and the background in the models of enhanced neutrino polarizability. As in the main text, we use the benchmark values for the neutrino polarizability and the mediator mass given in  \cref{eq:coupling:benchmark}. 

\begin{figure}[t]
	\centering
	\includegraphics[width=0.44\linewidth]{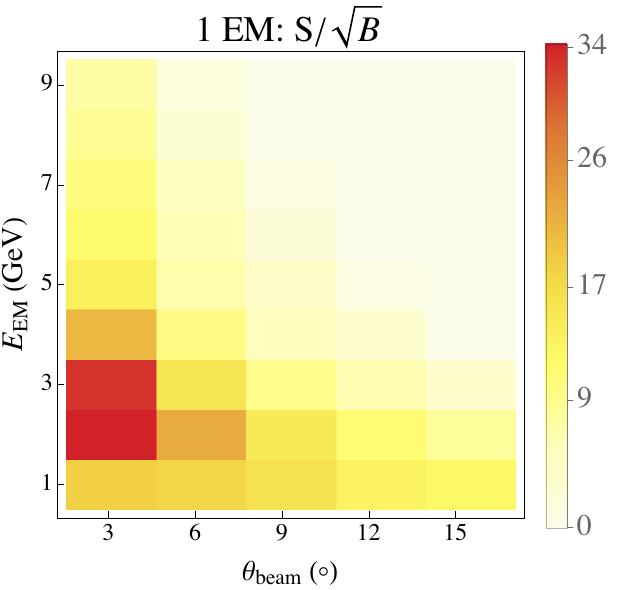}
    \hspace{1cm}
    \includegraphics[width=0.46\linewidth]{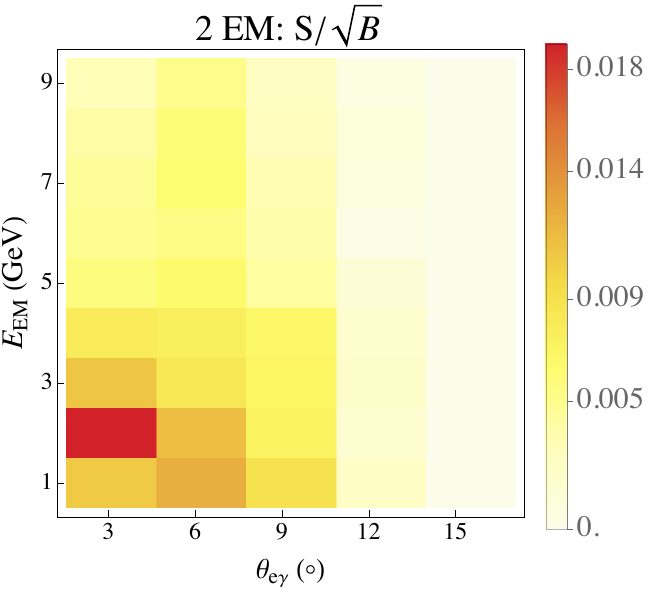}
	\caption{2D distribution of the $S/\sqrt{B}$ ratio for the 1EM (left) and 2EM (right) signature, respectively. }
	\label{fig:2DDist:SoverSqrtB}
\end{figure}

\Cref{fig:2DDist:SoverSqrtB} shows the values of the $S/\sqrt{B}$ ratio for each bin in the $(\theta_\text{beam}, E_\text{EM})$ plane for the 1EM signature (left panel), and in the $(\theta_{e\gamma}, E_\text{EM})$ plane for the 2EM signature (right panel). The binning and the assumed event rates for the signal and the background after 1 year exposure are the same as in \cref{fig:SvsB:1EM} and in \cref{fig:SvsB:2EM}, respectively. The results in \cref{fig:2DDist:SoverSqrtB} show that, were one able to ignore systematics, the effect of enhanced neutrino polarizability in the 1EM signature  would be highly visible for the benchmark value, while it would be still imperceptible in the 2EM signature analysis. 

\begin{figure}[t]
	\centering
	\includegraphics[width=0.64\linewidth]{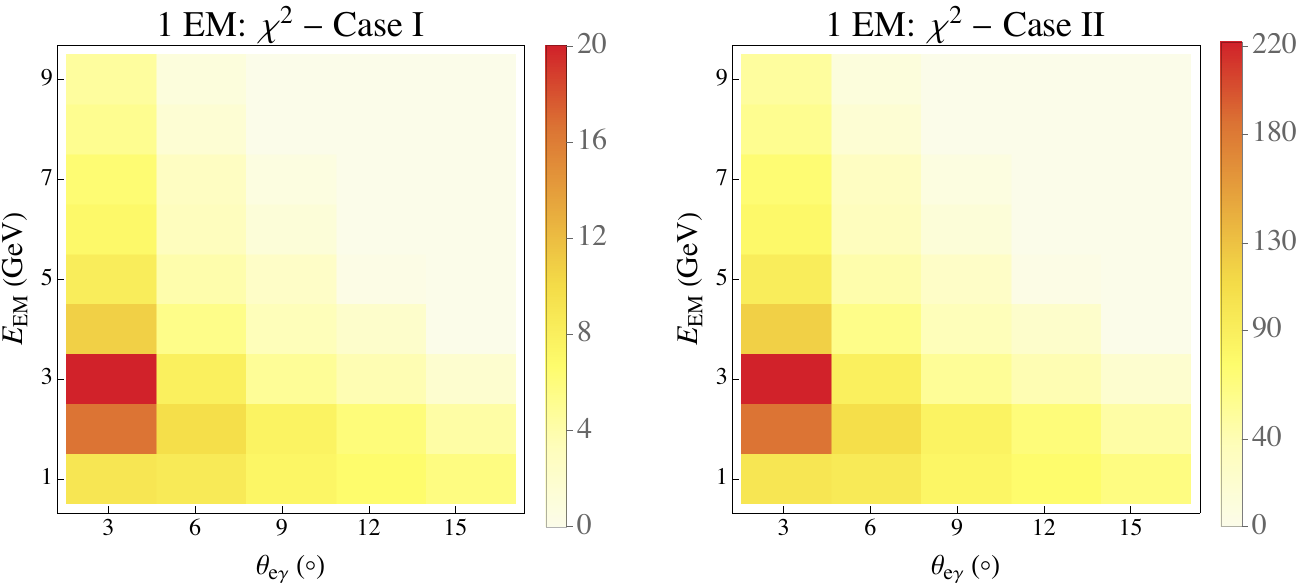} 
    \includegraphics[width=0.33\linewidth]{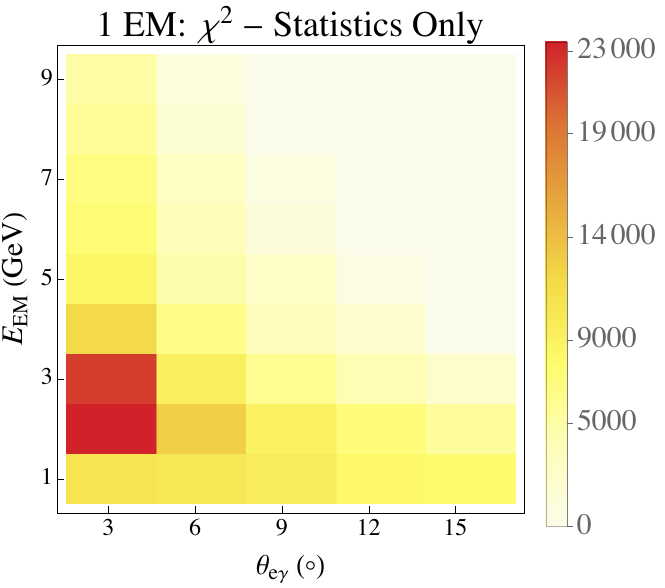}
	\caption{The 2D distributions of the $\chi^2$ values in each bin for the case of the 1EM signature, for the three statistics+systematics scenarios defined in \cref{eq.CaseI,eq.CaseII,eq.Stat:only}, as indicated on each panel.}
	\label{fig:2DDist:chi2:1EM}
\end{figure}

\begin{figure}[t]
	\centering
	\includegraphics[width=0.67\linewidth]{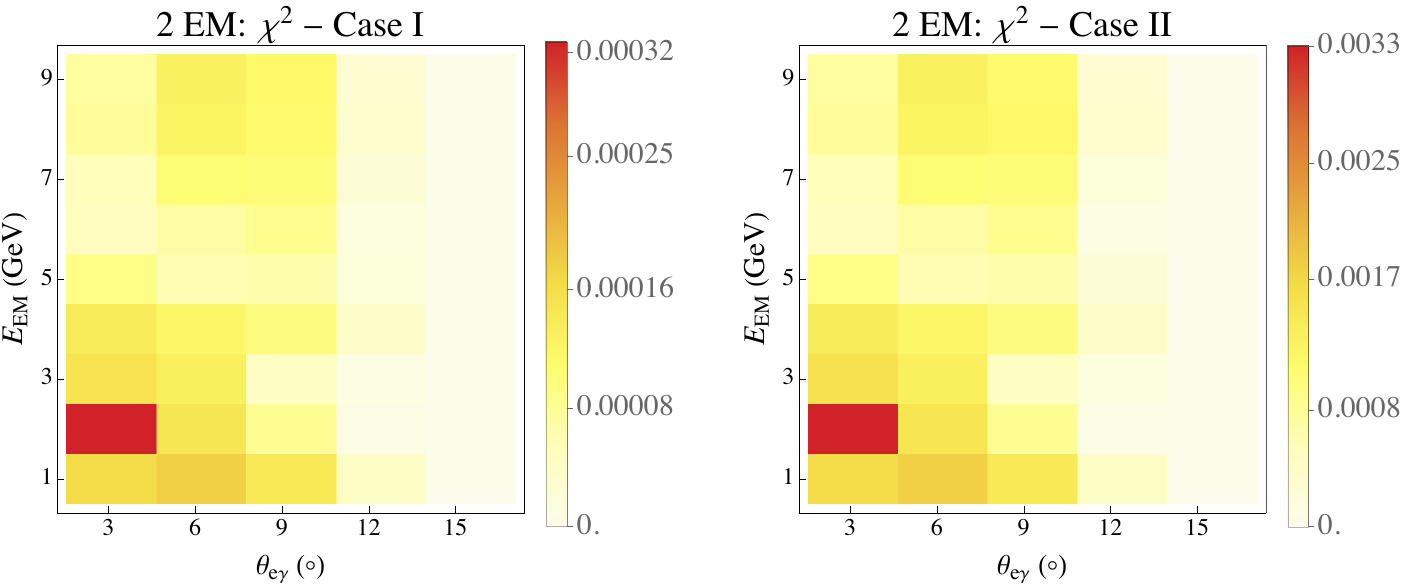} 
    \includegraphics[width=0.32\linewidth]{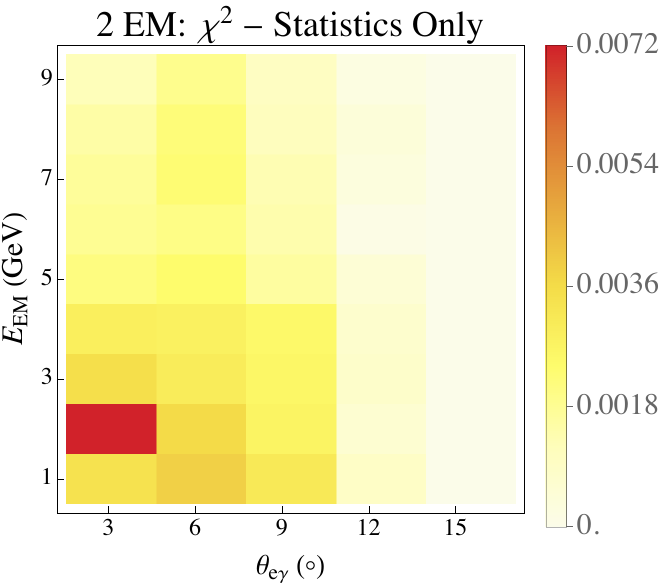}
	\caption{Same as \cref{fig:2DDist:chi2:1EM}, but for the 2EM signature.}
	\label{fig:2DDist:chi2:2EM}
\end{figure}

In \cref{fig:2DDist:chi2:1EM,fig:2DDist:chi2:2EM}, we show the values of $\chi^2$ distributions for each bin in the same angle vs. energy planes, assuming the three cases defined in \cref{eq.CaseI,eq.CaseII,eq.Stat:only}, for 1EM and 2EM signatures, respectively. The figures illustrate the increasing significance of the signal, with the increasing statistics and with reduced values of the systematics errors. 

\end{appendix}

%
%

\bibliographystyle{JHEP}
\bibliography{bib_nu_pol_DUNE}

\providecommand{\href}[2]{#2}\begingroup\raggedright\begin{thebibliography}{10}

\bibitem{Giunti:2024gec}
C.~Giunti, K.~Kouzakov, Y.-F. Li and A.~Studenikin, \emph{Neutrino
  electromagnetic properties},
  \href{http://dx.doi.org/https://doi.org/10.1146/annurev-nucl-102122-023242}{\emph{Annual
  Review of Nuclear and Particle Science} (2025) }.

\bibitem{Bansal:2022zpi}
S.~Bansal, G.~Paz, A.~Petrov, M.~Tammaro and J.~Zupan, \emph{{Enhanced neutrino
  polarizability}},
  \href{http://dx.doi.org/10.1007/JHEP05(2023)142}{\emph{JHEP} {\bf 05} (2023)
  142}, [\href{http://arxiv.org/abs/2210.05706}{{\tt 2210.05706}}].

\bibitem{BOREXINO:2018ohr}
{\scshape BOREXINO} collaboration, M.~Agostini et~al., \emph{{Comprehensive
  measurement of $pp$-chain solar neutrinos}},
  \href{http://dx.doi.org/10.1038/s41586-018-0624-y}{\emph{Nature} {\bf 562}
  (2018) 505--510}.

\bibitem{XENON:2020rca}
{\scshape XENON} collaboration, E.~Aprile et~al., \emph{{Excess electronic
  recoil events in XENON1T}},
  \href{http://dx.doi.org/10.1103/PhysRevD.102.072004}{\emph{Phys. Rev. D} {\bf
  102} (2020) 072004}, [\href{http://arxiv.org/abs/2006.09721}{{\tt
  2006.09721}}].

\bibitem{Giunti:2014ixa}
C.~Giunti and A.~Studenikin, \emph{{Neutrino electromagnetic interactions: a
  window to new physics}},
  \href{http://dx.doi.org/10.1103/RevModPhys.87.531}{\emph{Rev. Mod. Phys.}
  {\bf 87} (2015) 531}, [\href{http://arxiv.org/abs/1403.6344}{{\tt
  1403.6344}}].

\bibitem{Voloshin:1987qy}
M.~B. Voloshin, \emph{{On Compatibility of Small Mass with Large Magnetic
  Moment of Neutrino}}, {\emph{Sov. J. Nucl. Phys.} {\bf 48} (1988) 512}.

\bibitem{Babu:1989wn}
K.~S. Babu and R.~N. Mohapatra, \emph{{Model for Large Transition Magnetic
  Moment of the $\nu_e$}},
  \href{http://dx.doi.org/10.1103/PhysRevLett.63.228}{\emph{Phys. Rev. Lett.}
  {\bf 63} (1989) 228}.

\bibitem{Babu:2020ivd}
K.~S. Babu, S.~Jana and M.~Lindner, \emph{{Large Neutrino Magnetic Moments in
  the Light of Recent Experiments}},
  \href{http://dx.doi.org/10.1007/JHEP10(2020)040}{\emph{JHEP} {\bf 10} (2020)
  040}, [\href{http://arxiv.org/abs/2007.04291}{{\tt 2007.04291}}].

\bibitem{Babu:1990wv}
K.~S. Babu and R.~N. Mohapatra, \emph{{Large transition magnetic moment of the
  neutrino from horizontal symmetry}},
  \href{http://dx.doi.org/10.1103/PhysRevD.42.3778}{\emph{Phys. Rev. D} {\bf
  42} (1990) 3778--3793}.

\bibitem{Leurer:1989hx}
M.~Leurer and N.~Marcus, \emph{{A Model for a Large Neutrino Magnetic
  Transition Moment and Naturally Small Mass}},
  \href{http://dx.doi.org/10.1016/0370-2693(90)90466-J}{\emph{Phys. Lett. B}
  {\bf 237} (1990) 81--87}.

\bibitem{Altmannshofer:2018xyo}
W.~Altmannshofer, M.~Tammaro and J.~Zupan, \emph{{Non-standard neutrino
  interactions and low energy experiments}},
  \href{http://dx.doi.org/10.1007/JHEP11(2021)113}{\emph{JHEP} {\bf 09} (2019)
  083}, [\href{http://arxiv.org/abs/1812.02778}{{\tt 1812.02778}}].

\bibitem{Dreiner:2008tw}
H.~K. Dreiner, H.~E. Haber and S.~P. Martin, \emph{{Two-component spinor
  techniques and Feynman rules for quantum field theory and supersymmetry}},
  \href{http://dx.doi.org/10.1016/j.physrep.2010.05.002}{\emph{Phys. Rept.}
  {\bf 494} (2010) 1--196}, [\href{http://arxiv.org/abs/0812.1594}{{\tt
  0812.1594}}].

\bibitem{Duda:2006uk}
G.~Duda, A.~Kemper and P.~Gondolo, \emph{{Model Independent Form Factors for
  Spin Independent Neutralino-Nucleon Scattering from Elastic Electron
  Scattering Data}},
  \href{http://dx.doi.org/10.1088/1475-7516/2007/04/012}{\emph{JCAP} {\bf 04}
  (2007) 012}, [\href{http://arxiv.org/abs/hep-ph/0608035}{{\tt
  hep-ph/0608035}}].

\bibitem{DUNE:2021cuw}
{\scshape DUNE} collaboration, B.~Abi et~al., \emph{{Experiment Simulation
  Configurations Approximating DUNE TDR}},
  \href{http://arxiv.org/abs/2103.04797}{{\tt 2103.04797}}.

\bibitem{DUNE:2020ypp}
{\scshape DUNE} collaboration, B.~Abi et~al., \emph{{Deep Underground Neutrino
  Experiment (DUNE), Far Detector Technical Design Report, Volume II: DUNE
  Physics}},  \href{http://arxiv.org/abs/2002.03005}{{\tt 2002.03005}}.

\bibitem{DUNE:2016ymp}
{\scshape DUNE} collaboration, T.~Alion et~al., \emph{{Experiment Simulation
  Configurations Used in DUNE CDR}},
  \href{http://arxiv.org/abs/1606.09550}{{\tt 1606.09550}}.

\bibitem{MicroBooNE:2025ntu}
{\scshape MicroBooNE} collaboration, P.~Abratenko et~al., \emph{{Inclusive
  Search for Anomalous Single-Photon Production in MicroBooNE}},
  \href{http://arxiv.org/abs/2502.06064}{{\tt 2502.06064}}.

\bibitem{MiniBooNE:2018esg}
{\scshape MiniBooNE} collaboration, A.~A. Aguilar-Arevalo et~al.,
  \emph{{Significant Excess of ElectronLike Events in the MiniBooNE
  Short-Baseline Neutrino Experiment}},
  \href{http://dx.doi.org/10.1103/PhysRevLett.121.221801}{\emph{Phys. Rev.
  Lett.} {\bf 121} (2018) 221801}, [\href{http://arxiv.org/abs/1805.12028}{{\tt
  1805.12028}}].

\bibitem{Alwall:2014hca}
J.~Alwall, R.~Frederix, S.~Frixione, V.~Hirschi, F.~Maltoni, O.~Mattelaer
  et~al., \emph{{The automated computation of tree-level and next-to-leading
  order differential cross sections, and their matching to parton shower
  simulations}}, \href{http://dx.doi.org/10.1007/JHEP07(2014)079}{\emph{JHEP}
  {\bf 07} (2014) 079}, [\href{http://arxiv.org/abs/1405.0301}{{\tt
  1405.0301}}].

\bibitem{Golan:2012rfa}
T.~Golan, J.~T. Sobczyk and J.~Zmuda, \emph{{NuWro: the Wroclaw Monte Carlo
  Generator of Neutrino Interactions}},
  \href{http://dx.doi.org/10.1016/j.nuclphysbps.2012.09.136}{\emph{Nucl. Phys.
  B Proc. Suppl.} {\bf 229-232} (2012) 499--499}.

\bibitem{MicroBooNE:sepangle}
V.~Bhelande, ``Conducting beyond the standard model searches in the microboone
  detector with machine learning.''
  \url{https://indico.fnal.gov/event/68479/contributions/319827/}, Fermilab New
  Perspectives, 2025.

\bibitem{XENON:2022ltv}
{\scshape XENON} collaboration, E.~Aprile et~al., \emph{{Search for New Physics
  in Electronic Recoil Data from XENONnT}},
  \href{http://dx.doi.org/10.1103/PhysRevLett.129.161805}{\emph{Phys. Rev.
  Lett.} {\bf 129} (2022) 161805}, [\href{http://arxiv.org/abs/2207.11330}{{\tt
  2207.11330}}].

\bibitem{BaBar:2017tiz}
{\scshape BaBar} collaboration, J.~P. Lees et~al., \emph{{Search for Invisible
  Decays of a Dark Photon Produced in ${e}^{+}{e}^{-}$ Collisions at BaBar}},
  \href{http://dx.doi.org/10.1103/PhysRevLett.119.131804}{\emph{Phys. Rev.
  Lett.} {\bf 119} (2017) 131804}, [\href{http://arxiv.org/abs/1702.03327}{{\tt
  1702.03327}}].

\bibitem{Blinov:2019gcj}
N.~Blinov, K.~J. Kelly, G.~Z. Krnjaic and S.~D. McDermott, \emph{{Constraining
  the Self-Interacting Neutrino Interpretation of the Hubble Tension}},
  \href{http://dx.doi.org/10.1103/PhysRevLett.123.191102}{\emph{Phys. Rev.
  Lett.} {\bf 123} (2019) 191102}, [\href{http://arxiv.org/abs/1905.02727}{{\tt
  1905.02727}}].

\bibitem{Lessa:2007up}
A.~P. Lessa and O.~L.~G. Peres, \emph{{Revising limits on neutrino-Majoron
  couplings}}, \href{http://dx.doi.org/10.1103/PhysRevD.75.094001}{\emph{Phys.
  Rev. D} {\bf 75} (2007) 094001},
  [\href{http://arxiv.org/abs/hep-ph/0701068}{{\tt hep-ph/0701068}}].

\bibitem{Pasquini:2015fjv}
P.~S. Pasquini and O.~L.~G. Peres, \emph{{Bounds on Neutrino-Scalar Yukawa
  Coupling}}, \href{http://dx.doi.org/10.1103/PhysRevD.93.053007}{\emph{Phys.
  Rev. D} {\bf 93} (2016) 053007}, [\href{http://arxiv.org/abs/1511.01811}{{\tt
  1511.01811}}].

\bibitem{Lucente_2020}
G.~Lucente, P.~Carenza, T.~Fischer, M.~Giannotti and A.~Mirizzi, \emph{Heavy
  axion-like particles and core-collapse supernovae: constraints and impact on
  the explosion mechanism},
  \href{http://dx.doi.org/10.1088/1475-7516/2020/12/008}{\emph{Journal of
  Cosmology and Astroparticle Physics} {\bf 2020} (Dec, 2020) 008–008}.

\bibitem{Caputo:2022rca}
A.~Caputo, G.~Raffelt and E.~Vitagliano, \emph{{Radiative transfer in stars by
  feebly interacting bosons}},
  \href{http://dx.doi.org/10.1088/1475-7516/2022/08/045}{\emph{JCAP} {\bf 08}
  (2022) 045}, [\href{http://arxiv.org/abs/2204.11862}{{\tt 2204.11862}}].

\bibitem{Caputo:2021rux}
A.~Caputo, G.~Raffelt and E.~Vitagliano, \emph{{Muonic boson limits: Supernova
  redux}}, \href{http://dx.doi.org/10.1103/PhysRevD.105.035022}{\emph{Phys.
  Rev. D} {\bf 105} (2022) 035022},
  [\href{http://arxiv.org/abs/2109.03244}{{\tt 2109.03244}}].

\bibitem{Gehrlein:2025tko}
J.~Gehrlein, I.~M. Shoemaker and A.~Thapa, \emph{{Monophotons at Neutrino
  Experiments from Neutrino Polarizability}},
  \href{http://arxiv.org/abs/2506.14881}{{\tt 2506.14881}}.

\bibitem{Dutta:2025fgz}
B.~Dutta, A.~Karthikeyan, D.~Kim, A.~Thompson and R.~G. Van~de Water,
  \emph{{Photon Excess from Dark Matter and Neutrino Scattering at MiniBooNE
  and MicroBooNE}},  \href{http://arxiv.org/abs/2504.08071}{{\tt 2504.08071}}.

\bibitem{Helm:1956zz}
R.~H. Helm, \emph{{Inelastic and Elastic Scattering of 187-Mev Electrons from
  Selected Even-Even Nuclei}},
  \href{http://dx.doi.org/10.1103/PhysRev.104.1466}{\emph{Phys.Rev.} {\bf 104}
  (1956) 1466--1475}.

\bibitem{Klein:1999}
S.~R. Klein and J.~Nystrand, \emph{{Exclusive vector meson production in
  relativistic heavy ion collisions}},
  \href{http://dx.doi.org/10.1103/PhysRevC.60.014903}{\emph{Phys. Rev. C} {\bf
  60} (1999) 014903}, [\href{http://arxiv.org/abs/hep-ph/9902259}{{\tt
  hep-ph/9902259}}].

\bibitem{Yang:2019pbx}
J.~Yang, J.~A. Hernandez and J.~Piekarewicz, \emph{{Electroweak probes of
  ground state densities}},
  \href{http://dx.doi.org/10.1103/PhysRevC.100.054301}{\emph{Phys. Rev. C} {\bf
  100} (2019) 054301}, [\href{http://arxiv.org/abs/1908.10939}{{\tt
  1908.10939}}].

\bibitem{VanDessel:2020epd}
N.~Van~Dessel, V.~Pandey, H.~Ray and N.~Jachowicz, \emph{{Cross Sections for
  Coherent Elastic and Inelastic Neutrino-Nucleus Scattering}},
  \href{http://dx.doi.org/10.3390/universe9050207}{\emph{Universe} {\bf 9}
  (2023) 207}, [\href{http://arxiv.org/abs/2007.03658}{{\tt 2007.03658}}].

\bibitem{Payne:2019wvy}
C.~G. Payne, S.~Bacca, G.~Hagen, W.~Jiang and T.~Papenbrock, \emph{{Coherent
  elastic neutrino-nucleus scattering on $^{40}$Ar from first principles}},
  \href{http://dx.doi.org/10.1103/PhysRevC.100.061304}{\emph{Phys. Rev. C} {\bf
  100} (2019) 061304}, [\href{http://arxiv.org/abs/1908.09739}{{\tt
  1908.09739}}].

\end{thebibliography}\endgroup

\end{document}